\documentclass{aa_noline}  

\usepackage{graphicx}
\usepackage{txfonts}

\usepackage{hyperref}

\usepackage{xcolor}
\usepackage{ulem}

\usepackage{amsmath}    

\def\nbR{\ensuremath{\mathrm{I\! R}}}
\def\nbE{\ensuremath{\mathrm{I\! E}}}

\newcommand{\R}{\textsf{R}}

\newcommand{\HII}{\ion{H}{II}}
\newcommand{\OIII}{\ion{O}{III}}
\newcommand{\trCO}{\element[][13]{CO}}

\begin{document} 

      \title{Spectral similarities in galaxies through an unsupervised classification of spaxels}   
\titlerunning{Unsupervised classification of spaxels}

   \author{
        H.J. Chambon\inst{1}
        \and
        D. Fraix-Burnet\inst{1}
   }
   
   \institute{Univ. Grenoble Alpes, CNRS, IPAG, Grenoble, France \\ \email{hugo.chambon@univ-grenoble-alpes.fr}
   }
   
   \date{Received January, 2024; accepted May, 2024}
   
\abstract
{}
{
We present the first unsupervised classification of spaxels in hyperspectral images of individual galaxies. Classes identify regions by spectral similarity and thus take all the information into account that is contained in the data cubes (spatial and spectral).
}
{
We used Gaussian mixture models in a latent discriminant subspace to find clusters of spaxels. The spectra were corrected for small-scale motions within the galaxy based on emission lines with an automatic algorithm. Our data consist of two MUSE/VLT data cubes of JKB~18 and NGC~1068 and one NIRSpec/JWST data cube of NGC~4151.
}
{
Our classes identify many regions that are most often easily interpreted. Most of the 11 classes that we find for JKB~18 are identified as photoionised by stars. Some of them are known \HII\ regions, but we mapped them as extended, with gradients of ionisation intensities. One compact structure has not been reported before, and according to diagnostic diagrams, it might be a planetary nebula or a denser \HII\ region. For NGC~1068, our 16 classes are of active galactic nucleus-type (AGN) or star-forming regions. Their spatial distribution corresponds perfectly to well-known structures such as spiral arms and a ring with giant molecular clouds. A subclassification in the nuclear region reveals several structures and gradients in the AGN spectra. Our unsupervised classification of the MUSE data of NGC~1068 helps visualise the complex interaction of the AGN and the jet with the interstellar medium in a single map. The centre of NGC~4151 is very complex, but our classes can easily be related to ionisation cones, the jet, or H$_2$ emission. We find a new elongated structure that is ionised by the AGN along the N-S axis perpendicular to the jet direction. It is rotated counterclockwise with respect to the axis of the H$_2$ emission. 

}
{
Our work shows that the unsupervised classification of spaxels takes full advantage of the richness of the information in the data cubes by presenting the spectral and spatial information in a combined and synthetic way. 
}

   \maketitle
   
   \section{Introduction}
   \label{section:introduction}

Hyperspectral imaging (or integral field spectroscopy; IFS thereafter) in astrophysics probably yields the most complete observations that might be imagined. Spatial and spectral information together are available in the same data set. This is particularly welcome in the case of galaxies, which are complex structures with a great variety of physical features and kinematic behaviour. IFS generates data cubes that are intrinsically computationally intensive to analyse. As numerous new instruments and dedicated surveys provide a huge number of these data cubes, new approaches are needed, and machine-learning can come to our rescue.

A single data cube contains so much information that it requires many graphs to visualise the properties of different regions. A first example that has been used by radio-astronomers for a long time is the representation of some absorption or emission lines along the line of sight using kinematics as the third dimension. This yields a sort of movie in this third dimension perpendicular to the plane of the sky. In the optical domain, it is possible to generate maps of line intensities, line ratios, and kinematics for the numerous lines present, along with diagnostic diagrams that are generally based on line ratios \citep[e.g.][]{DAgostino2019,Venturi2021}. Interpreting and synthesising these many 3D maps is a challenging task. It is time-consuming even for a single object, so that automatic tools should be used.

 In all cases, line properties are computed individually for each spectrum given by each spaxel. In the optical domain, the high resolution and the complexity of galaxies most often result in a low signal-to-noise ratio per spaxel. In addition, the scale of individual spaxels can become smaller than the typical size of \ion{H}{II} regions or the largest molecular clouds for which the methods for deriving some physical quantities have been established \citep[see a discussion in ][]{Bulichi2023}. A common workaround is to bin several close spaxels together, which effectively results in a decrease of the spatial resolution. The reasoning behind this workaround is that two spaxels that are close in space should also be close in physical terms. This can be seen as a clustering of spaxels based on spatial similarity, with the spatial size being guided by a criterion that is the signal-to-noise ratio.

In the present work, we propose a more physical approach by grouping spaxels based on spectral similarity using unsupervised classification under objective statistical criteria. Each class gathers similar spectra and is thus characterised by a mean or median spectrum with a higher signal-to-noise ratio. The within-class dispersion is also lower because different physical regions are not mixed. Consequently, the original resolution and the spatial sampling are preserved, and gradients and boundaries remain as they are in the image. This allows the detection of truly coherent structures. In addition to this benefit from the physical point of view, unsupervised classification of spaxels provides an automatic technique for synthesising the data cube into a unique map containing both the spatial and multivariate spectral information.
Because it is unsupervised and takes the entire spectra, our approach could lead to the discovery of new structures.

To our knowledge, only two very recent works in astrophysics proposed to gather spaxels based on their spectral similarity using unsupervised classification. 
\citet{Rosito2023} considered the unsupervised classification of the kinematic morphologies of simulated galaxies. Their automatic method is able to achieve not only a clear division between slow and fast rotators, but also a segregation in the rotation orientation and subdivisions among low-rotation edge-on galaxies corresponding to their shapes. However, kinematic maps are only a small part of the information contained in IFS data cubes. \citet{Tiwari2023} applied a slightly modified version of the Gaussian mixture model (GMM) clustering technique to spectra of 50 channels (wavelengths) around the [CII] 158 $\mu$m line to identify coherent physical structures in the interstellar medium in our Galaxy. In each of the three Galactic sources they studied, they found a few such structures. This compares favourably with the literature.

These two studies validated the use of automatic unsupervised classification tools. In the present paper, we extend these results to full optical spectra. It is crucial to take into account all the information contained in the spectra to derive the detailed characteristics of different regions in a galaxy. In particular, diagnostic diagrams are most often 2D only, and several diagrams are needed to disentangle degeneracies \citep[e.g.][]{Johnston2023}. 

We therefore apply an unsupervised analysis on data cubes for individual galaxies. The spectra can have several thousand wavelengths (channels), which means that the problem has a high dimensionality. As a consequence, we apply a GMM technique in a latent discriminant subspace that is optimised for the clustering. We make use of the algorithm Fisher-EM \citep{Bouveyron2012}, which has been used several times in astrophysics \citep{Siudek2018,Siudek2018a,Fraix-Burnet2021,Siudek2022,Dubois2022,Fraix-Burnet2023,Dubois2024}. 
We showed in \citet{Fraix-Burnet2021} that the clustering of galaxy spectra with the Fisher-EM algorithm is able to distinguish according to the shape of the continuum, to the intensity and shape of the emission and absorption lines, and to the line ratios. This means that the classes of spectra include all the 2D or 3D maps that are generally used to understand the data cubes, as mentioned earlier. In other words, classes represent coherent structures that are identical in a multivariate space and not only in some peculiar property.

In this paper, we demonstrate the capabilities of an unsupervised classification of spaxels on two galaxies that were observed with the multi-unit spectroscopic explorer installed at the very large telescope (MUSE/VLT) and one galaxy that was observed with the near infrared spectrograph on the James Webb space telescope (NIRSpec/JWST). 

\object{JKB~18} is a nearby (z = 0.004) blue diffuse dwarf galaxy. Blue diffuse galaxies are a population of low-metallicity dwarf galaxies that form stars \citep{James2016,James2020}. The distance and mass of JKB 18 are estimated as $\sim$18~Mpc and  $\sim$$10^{8}$~M$_\sun$. Its morphology shows numerous sites of ongoing star formation spread over a diffuse body. This type of object is considered a nearby analogue of the first galaxies, allowing us to study the spatial distribution of metals and the chemical homogeneity of these currently unobservable systems. JKB~18 is an ideal target for a first unsupervised classification analysis of IFS observations.

\object{NGC~1068} is a prototypical Seyfert 2 galaxy located at a distance of $\sim$10.5~Mpc. It is one of the brightest Seyfert galaxies at radio wavelengths. It hosts a radio jet spanning up to $\sim$800 pc in the NE-SW direction, and its powerful starburst activity is mainly concentrated in a prominent starburst ring with a radius of $\sim$1-1.5~kpc \citep{Venturi2021}. In addition, the gas in the disc of the galaxy is illuminated by the AGN radiation and interacts with the jet. In the field of view of the MUSE instrument of $\sim$3.3~kpc, many regions with various spectral characteristics are found, which makes this target a particularly rich case study for an unsupervised classification analysis.

\object{NGC~4151} is the brightest Seyfert 1 galaxy in the sky and situated at a distance of 18.5~Mpc. It has a very complex central region: The line of sight seems to be close to the edge of a clumpy torus, the ionisation cones show a highly filamentary [O III] emission that is not directly associated with the radio jet, and some localised regions have peculiar line properties and even some indications of shocks \citep[see an overview in][]{May_2020}. With the very high spatial resolution of the NIRSpec instrument on board JWST, it represents a challenge for our goal, but it might illustrate how an automatic unsupervised classification of spaxels could synthesise very complex data cubes.

This paper is organised as follows. The data are presented in Sect.~\ref{Data}. The algorithm Fisher-EM is described in Sect.~\ref{Method}. Our results and discussion are gathered in separate sections for each object: Sect.\ref{JKB18} for JKB~18, Sect.~\ref{NGC1068} for NGC~1068, and Sect.~\ref{NGC4151} for NGC~4151.
We conclude this paper in Sect.~\ref{Conclusion}.

\section{Data}
\label{Data}

\subsection{MUSE}
\label{MUSE}

The field of view of MUSE of one arcminute is covered by 319x319 spaxels ($\sim$100,000 spectra) of 0.2~arcsec each. 

\subsubsection{JKB~18}
\label{Data:JKB18}

The MUSE observation of JKB~18 were thoroughly studied by \citet{James2020}, and we take this work as a reference. 

We took the scientific data from the ESO archive\footnote{http://archive.eso.org/cms.html} (ADP.2016-09-06T09:08:42.261) without reprocessing the data as \citet{James2020} did, who tried to subtract the sky better.

We used the same automated line-fitting algorithm \citep[ALFA, ][]{10.1093/mnras/stv2946} to obtain the kinematics and line properties. Our kinematics was not only determined from the H$\alpha$ line, however, but rather from a catalogue of several strong lines in the range [6400, 6800]~\AA: the \ion{N}{II} doublet at 6548,6583~\AA, H$_\alpha$ at 6584~\AA, and the \ion{S}{II} doublet at 6716, 6731~\AA. As the unsupervised classification requires all spectra to be rigorously aligned for them to be compared, we corrected them all to be at the same redshift zero. 

Spaxels for which ALFA returned a discrepant or unreliable redshift were discarded. This included the bright foreground star and most of the sky.  The frame border additionally induces distortion in the spectra, and the data outside the spaxel range [52, 290] were discarded. We finally analysed 19516 spaxels.

\citet{James2020} based the identification of the \ion{H}{II} regions and their sizes on the H$\alpha$ line through a compromise between the number of \ion{H}{II} regions and a sufficient signal-to-noise ratio. In our work, the unsupervised algorithm distinguishes regions with similar spectra without any a priori assumption on their physical nature.

Finally, \citet{James2020} derived the chemical abundances from the spectra of individual spaxels or integrated over each \ion{H}{II} region. In our work, they are derived from the median spectrum of each class.

\subsubsection{NGC~1068}
\label{Data:NGC1068}

The scientific data were retrieved from the  european southern observatory (ESO) archive (ADP.2016-06-17T08:44:56.817) without further reprocessing of the sky. We removed the edges of the hyperspectral image and kept the spaxels in the range [25,295]. 

Because the continuum levels of the different regions of NGC~1068 are much broader than those of JKB~18, we normalised the spectra to their mean within the range [7600, 8100]~\AA\ to avoid a classification mainly dictated by intensity.

We used the ALFA algorithm to obtain the kinematics and line properties with the same catalogue of emission lines. However, the complex kinematics of the nucleus caused us to consider exploring a broader range of velocities for ALFA ($\pm$ 2000 $km\;s^{-1}$). This increased the risk of falling into local minima, which would yield an incorrect redshift estimate. Spaxels based on an estimation of unreliable redshift were removed. This resulted in 68232 spaxels for the unsupervised classification.

For NGC~1068, we also performed a clustering analysis without aligning the spectra, that is, without correcting for the kinematics, in order to test whether an unsupervised analysis can take the spectral shift due to the internal motions within a galaxy into account, in addition to all the intrinsic spectral components.

\subsection{NIRSpec}
\label{NIRSpec}

The scientific observation of the nucleus of NGC~4151 was retrieved from the Mikulski Archive for Space Telescopes\footnote{https://mast.stsci.edu/portal\_jwst/Mashup/Clients/Mast/Portal.html} (jw01364-o001\_t001\_nirspec\_g235h-f170lp\_s3d).

This observation focuses on the very centre of the nucleus, with a field of view of 3~arcsec ($\sim$228~pc). The number of spaxels is about 1000, and the filter used (F170LP;G235H) imposes a gap between 2.36 and 2.491~$\mu m$ \citep{refId0} that we discarded for the clustering analysis. The spectra  were normalised using the range [20000, 21000]~\AA. 

We used the ALFA algorithm to obtain the kinematics and line properties with a strong line catalogue adapted to the near-infrared range: the Pa$_\alpha$ line at 18748, \AA, the [\ion{Si}{VI}] line at 19634 \AA, the Br$_\beta$ at 26258 \AA, and the [\ion{Mg}{VIII}] line at 30279~\AA\ . Owing to the extreme kinematics inside the nucleus of NGC~4151, we used a velocity exploration range of $\pm$ 1500 $km\;s^{-1}$.
The spectra of the spaxels where the kinematics search failed were removed.

\section{Method: Fisher-EM algorithm}
    \label{Method}

\subsection{Gaussian mixtures models in a latent subspace}

Clustering divides a given data set $Y=\{ y_{1},...,y_{n}\}$ of $n$
data points into $K$ homogeneous groups. Here, our data were the set of spectra given by the spaxels of the data cube: $y_{i \in [1..n]} \in \mathbb{R}^{p}$ where the dimension $p$ is the $p$ fluxes at the $p$ wavelengths of the spectra. A popular clustering technique uses GMM, which assumes that each class is represented by a Gaussian
probability density. The data are therefore modelled by a density, 
\begin{equation}
f(y,\theta)=\sum_{k=1}^{K}\pi_{k}\phi(y,\theta_{k}),
\label{eqHDDC:mixture-model}
\end{equation}
where $\phi$ is a $p$-variate normal density, in which the parameter $\theta_{k}=\{\mu_{k},\Sigma_{k}\}$ contains the means $\mu_{k}$ and the covariance matrices $\Sigma_{k}$, and $\pi_{k}$ are the mixing proportions. This model requires estimating full covariance matrices $\Sigma_{k}$, and therefore, the number of parameters increases with the square of the dimension. 
The number of parameters that is to be estimated can be limited by assuming that high-dimensional data lie in subspaces with a dimension lower than that of the original space \citep[e.g.][]{Bouveyron2019}.

The Fisher-EM algorithm fits the data in a low-dimensional optimised subspace that is common to all clusters. This simplifies their comparison. 
This algorithm is available in the \R\footnote{\url{https://www.r-project.org/}}\ package called FisherEM \citep[command \textit{fem},][]{Bouveyron2012}.

In Appendix~\ref{Ap:method:HDDC}, we present a similar algorithm, called high-dimensional data clustering (HDDC), in which the latent subspaces are class-specific. We used it here only in the difficult case of NGC~4151 for the comparison with the result of the Fisher-EM algorithm.

   \subsection{Fisher-EM algorithm}
   \label{method:Fisher-EM} 
   
   Fisher-EM \citep{Bouveyron2012} is a discriminant latent subspace Gaussian mixture algorithm. It uses a modified version of the expectation-maximisation (EM) algorithm by inserting a Fisher-step that optimises the ratio of the sum of the between-class variance over the sum of the within-class variance, hence optimising the Gaussian mixture and the subspace together for a better clustering.
   
  Classifying the observations into K classes mathematically translates into finding the vector ${Z}=\{z_1, ..., z_n\}$, which assigns each spectrum $y_i$ to a given class $z_i \in  [\![1,K]\!]$. 
  In the case of Fisher-EM, the clustering process occurs in a subspace $\nbE \subset \nbR^p$ of dimension $d=K-1 < p$. 
Therefore, the GMM is applied to the projected data ${X}$ rather than the observed data ${Y}$:
   \begin{equation}
   \centering
       {Y} = {U}{X} + {\epsilon}
       \label{eq:Y},
   \end{equation}
   where ${U} \in \mathcal{M}_{p,d}(\nbR)$ is the projection matrix, and ${\epsilon}$ is a noise vector of dimension $p$ following a Gaussian distribution centred around 0 and of the covariance matrix $\Psi$ ($\varepsilon_{k}\sim\mathcal{N}(0,\Psi_{k})$). Hence,
   \begin{equation}
       \centering
       X|_{Z=k} \sim \mathcal{N}(\mu_k, \Sigma_k)
       \label{eq:X}.
   \end{equation}
   Combining Eqs.~\ref{eq:Y} and \ref{eq:X}, we obtain
   \begin{equation}
       \centering
       Y_{|X,Z=k}\sim\mathcal{N}(UX,\Psi_{k}).
   \end{equation}
   The observed data are thus modelled by a marginal distribution $f({y})$ that is the sum of K multivariate Gaussian density functions $\phi$ of mean ${U}\mu_k$ and the covariance ${U}\Sigma_k{U^t} + \Psi$, each weighted by the corresponding mixing proportion $\pi_k$,
    \begin{equation}
       \centering
       f({y}) = \sum_{k=1}^K \pi_k \phi(y;{U}\mu_k, {U}\Sigma_k{U^t} + \Psi)
\label{eq:f}.
   \end{equation}
   
   By further assuming that the noise covariance matrix $\Psi_{k}$ 
satisfies the conditions $V^{t}\Psi_{k} V=\beta_{k}\mathbf{I}_{p-d}$, where $V$ is the orthogonal complement of $U$, and $U^{t}\Psi_{k} U=\mathbf{0}_{d}$, the whole statistical model denoted by $\mathrm{DLM}_{[\Sigma_{k}\beta_{k}]}$
can be shown to take the following form: $$ \Delta_k=\left(  \begin{array}{c@{}c} \begin{array}{|ccc|}\hline ~~ & ~~ & ~~ \\  & \Sigma_k &  \\  & & \\ \hline \end{array} & \mathbf{0}\\ \mathbf{0} &  \begin{array}{|cccc|}\hline \beta_{k} & & & 0\\ & \ddots & &\\  & & \ddots &\\ 0 & & & \beta_{k}\\ \hline \end{array} \end{array}\right)  \begin{array}{cc} \left.\begin{array}{c} \\\\\\\end{array} \right\}  & d \leq K-1\vspace{1.5ex}\\ \left.\begin{array}{c} \\\\\\\\\end{array}\right\}  & (p-d).\end{array}$$ These
last conditions imply that the discriminative and the non-discriminative
subspaces are orthogonal, which suggests in practice that all the
relevant clustering information remains in the latent subspace. From a practical point of view, $\beta_{k}$ models the variance of the non-discriminative noise of the data.
   
   Several other models can be obtained from the DLM$_{[\Sigma_{k}\beta_{k}]}$
model by relaxing or adding constraints on the model parameters. For example, it can be assumed that the noise parameter $\beta_{k}$ differs from one class to another, or that the covariance matrices $\Sigma_k$ are the same for all $K$ classes.
   A thorough description of the DLM models with its 12 declinations can be found in \cite{Bouveyron2012}.

The estimation procedure, called the Fisher-EM algorithm, is used to estimate the discriminative
space and the parameters of the mixture model (including the best statistical DLM model and the optimum number of clusters). This algorithm is based
on the EM algorithm, to which an additional step is introduced between
steps E and M. This additional step, called step F, aims to
compute the projection matrix $U$ whose columns span the discriminative
latent space. In Fisher-EM, $d=K-1$ is imposed.

The choice of the best statistical DLM model and the optimum number of clusters depends on the data. We estimated them with the integrated completed likelihood (ICL) criterion. This criterion penalises the likelihood by the number of observations, the number of parameters of the statistical model, and favours well-separated clusters \citep{Biernacki2000,Girard2016}. It generally has nearly identical values as the well-known criterion Bayesian information criterion (BIC). It is a purely objective criterion that only relies on statistical arguments without any tunable parameter or hyper-parameter. Its maximum value corresponds to the best set of the Gaussian models, that is, to the number of clusters, the parameters of the corresponding Gaussian distributions, and the subspace (Eq.~\ref{eq:f}). The solution must be interpreted from the physical point of view and validated if the clusters are found to have coherent and consistent physical properties.

\subsection{Subclassification}
\label{method:subclassification}

As already stated, the choice of the best statistical model and number of clusters depend on the data. In particular, the discriminant subspace depends on the distribution of all the data points. This implies that the optimum subspace may differ when only a subset (e.g. a class) is considered. A different subspace potentially leads to different structures.
This is called subclassification.
In the case of spectra, it is easy to imagine that the first classification would separate broad categories depending on the presence or absence of emission lines, and subsequent subclassifications would be more sensitive to faint lines or line ratios. This was well illustrated in our previous analyses \citep{Fraix-Burnet2021,Dubois2024}. 

There is no absolute rule to decide which class should be subclassified. It entirely depends on the characteristics of the class, such as the fraction of the sample it contains or the heterogeneity (dispersion) of its spectra. The subclassification can also be repeated several times to obtain more granularity, but this entirely depends on the goal of the study. Most often, however, the algorithm Fisher-EM itself will provide some clue, for instance when it fails to find a solution or yields no significantly improved result, such as when it yields a large subclass containing nearly all spectra and other very small subclasses.

\section{Analysis of JKB~18}
\label{JKB18}

\subsection{Results}
\label{JKB18_results}

\begin{figure}
                \includegraphics[width=\linewidth]{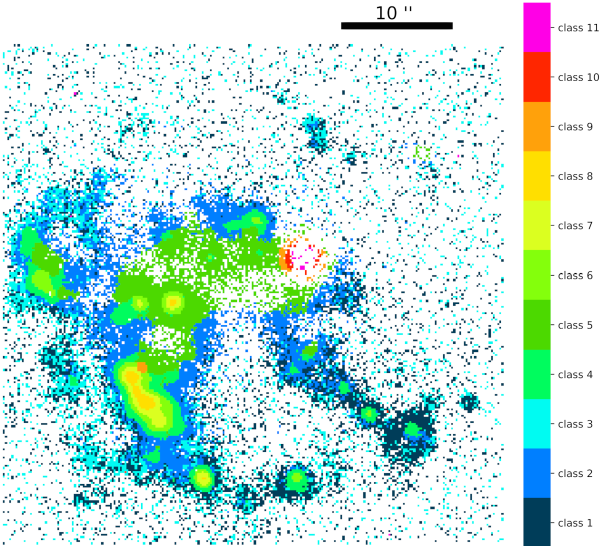}
                \caption{Class map for JKB~18. The orange, red, and violet rings near the centre of the image are due to the residual of a bright foreground star that was removed by discarding spectra with a redshift of zero. In all the class maps in this paper, the order of the class numbers is chosen to follow the similarity of the mean spectra of the classes as determined by the Euclidean distance, and white pixels are discarded spaxels for which ALFA returned a discrepant or unreliable redshift .}
                \label{Fig:JKB18classmap}
\end{figure}

\begin{figure}
        \includegraphics[width=\linewidth]{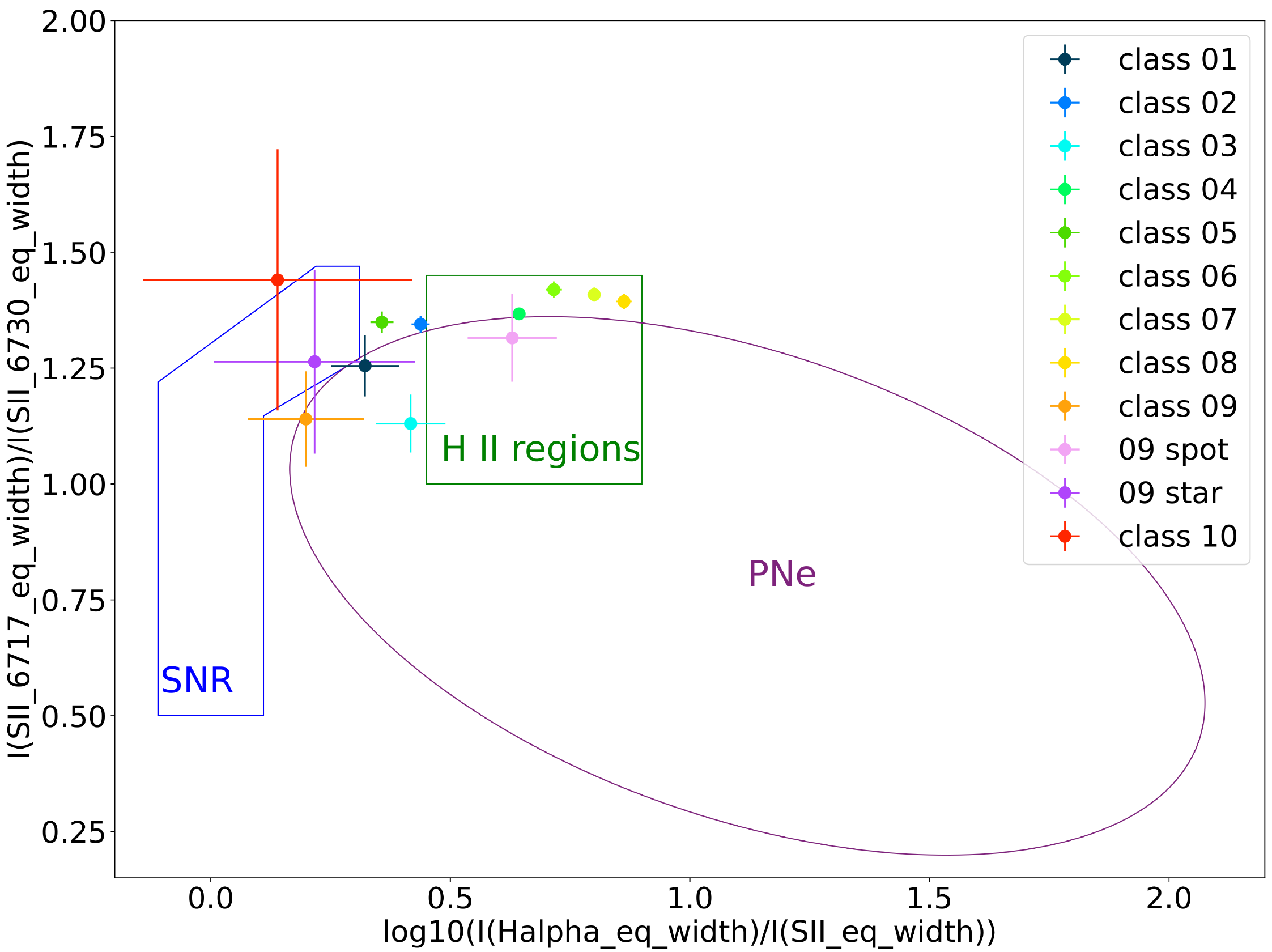}
        \caption{Electron density-excitation diagram for JKB~18. The contoured regions are taken from \citet{Riesgo2006}. The error bars are derived from line-fitting uncertainties given by ALFA.}
        \label{Fig:JKB18Polcaro}
\end{figure}

\begin{figure}
        \includegraphics[width=\linewidth]{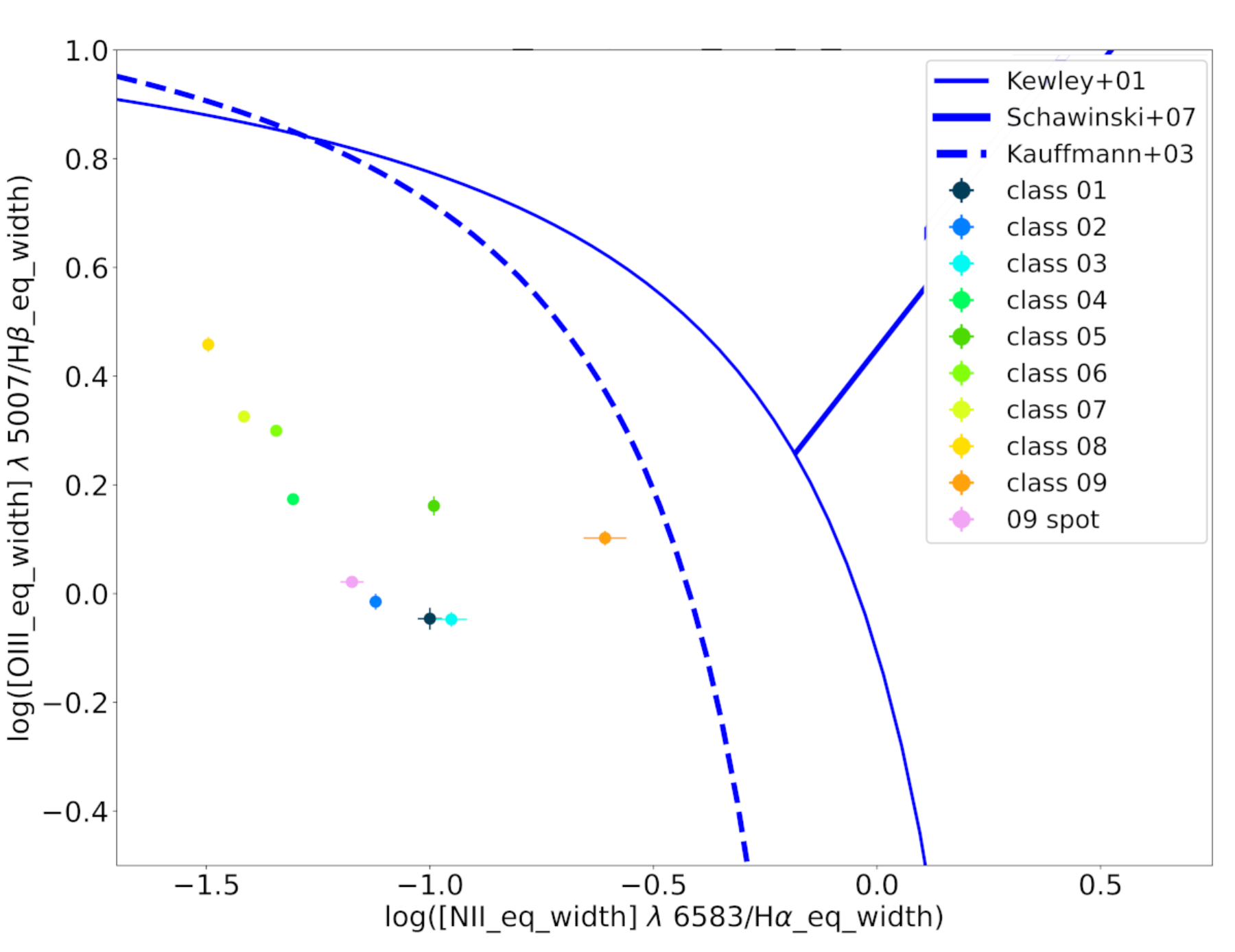}
        \caption{BPT diagram for JKB~18. The error bars are the same as in Fig.~\ref{Fig:JKB18Polcaro}.}
        \label{Fig:JKB18BPT}
\end{figure}

\begin{figure}
        \includegraphics[width=\linewidth]{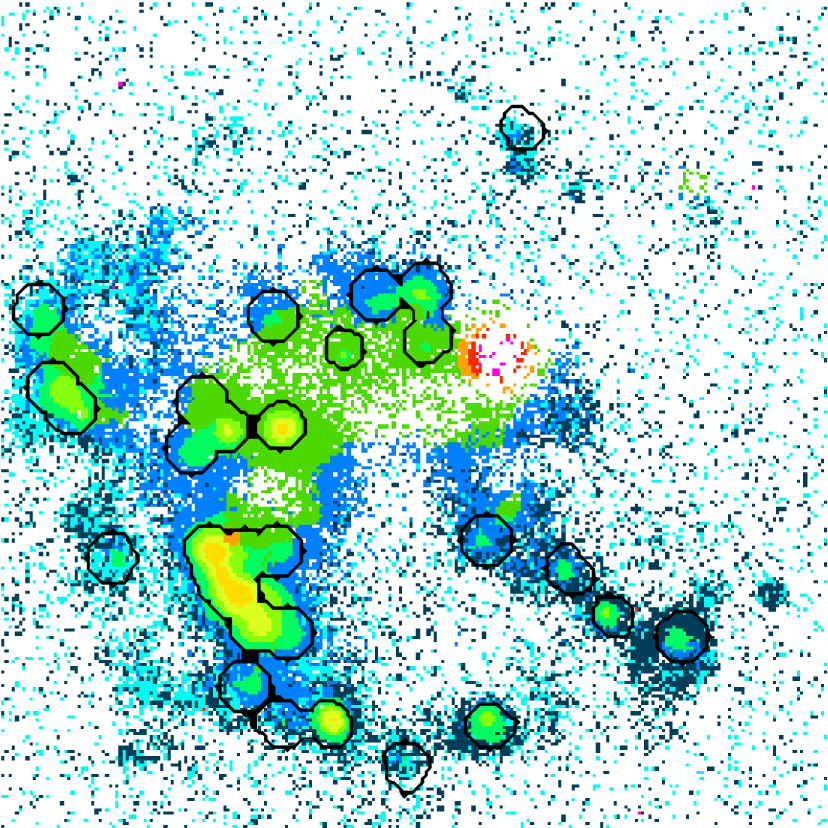}
        \caption{Class map for JKB~18 with contours of \ion{H}{II} regions identified by \citet{James2020} overlaid.}
        \label{Fig:JKB18classmap_sup}
\end{figure}

\begin{figure}
        \includegraphics[width=\linewidth]{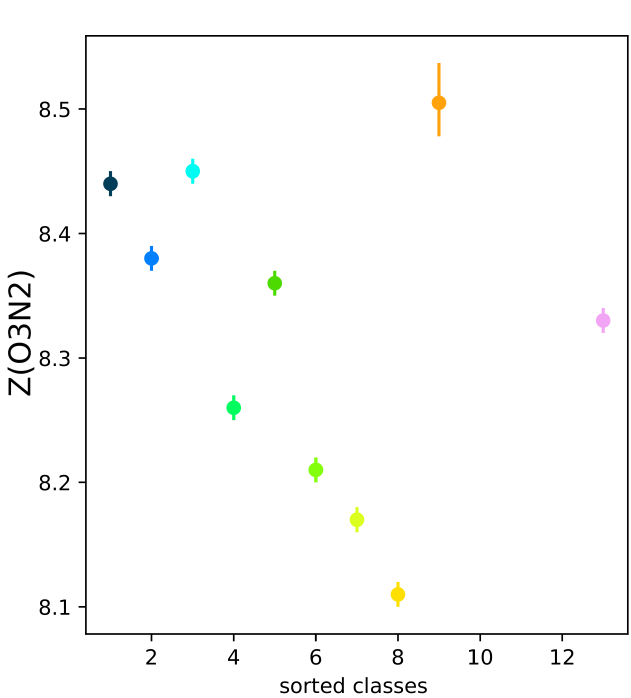}
        \caption{Metallicity for the classes for JKB~18. The values and error bars were computed through the algorithm NEAT \citep{2012MNRAS.422.3516W}}
        \label{Fig:JKB18O3N2}
\end{figure}

An optimum of 11 classes was found for JKB~18 (Fig.~\ref{Fig:JKB18classmap}).
Many structures are detected, with round and concentrated structures in the bottom right quadrant of the image and extended structures in the bottom left quadrant. Most compact regions show a gradient of classes, the most conspicuous structures having always the same order of the classes 4, 6, 7, and 8 inwards. This corresponds to increasing emission line intensities (Fig.~\ref{Fig:JKB18_AllMedSpec}).

The ring-like structure just above and to the right of the centre of the image is a residual of a bright foreground star. It is characterised by three classes, two inner classes (classes 10 and 11) that are not considered any further in the following, and an outer class (class 9). Class 9 is also found in a small round feature in the bottom left quadrant that we call "the spot" in the rest of this paper. It appears that the median spectra of the spot and the rest of class 9 (Fig.~\ref{Fig:JKB18_AllMedSpec}) share a common property that is specific to this class: The continuum is higher than in the other classes. This explains why Fisher-EM placed all these spectra in the same class even though emission lines are only present in the spot spectrum. This particularity and the zone around the foreground star that generates artefacts justifies a subclassification analysis of this class 9. It is classified into four subclasses that almost perfectly discriminate the spot.

\subsection{Discussion}
\label{JKB18_discussion}

Diagnostic diagrams are useful tools for identifying the source of ionisation. The electron density-excitation diagrams compare the relative line intensity ratios observed in planetary nebulae, supernova remnants, and \HII\ regions \citep{Sabbadin1977,Riesgo2006}, that is, of sources that are ionised by stars. We used the [\ion{S}{II}]\ 6717/[\ion{S}{II}]\ 6731, and H$_\alpha$/[\ion{S}{II}] ratio diagram (Fig.~\ref{Fig:JKB18Polcaro}). Classes 1 and 3 lie in the region of planetary nebulae. Classes 2 and 5 seem outside any region, but very close to planetary nebulae or \ion{H}{II} for the former and very close to a supernova remnant or \ion{H}{II} for the latter. The four classes 4, 6, 7, and 8 lie in \ion{H}{II} regions, and the inward gradient is horizontal in this diagram. The spot (of class 9) lies in the overlapping zone between the \ion{H}{II} and the planetary nebula regions. Classes 9 (star) and 10 lie in the supernova remnant region, but this is certainly not significant because they are artefacts caused by the foreground star.

The Baldwin-Phillips-Terlevich (BPT) diagram \citep{Baldwin1981} has been designed to identify ionising sources in galaxies between star-forming regions (photoionised by O and B stars), shock-heating, or non-thermal photoionisation. Planetary nebulae and \ion{H}{II} regions are included in the star-forming region because they are generally difficult to isolate in external galaxies. The BPT diagram for our JKB~18 classes (Fig.~\ref{Fig:JKB18BPT}) shows that all classes are in the star-forming zone. The four classes 4, 6, 7, and 8 are well separated from the other classes in the upper left corner, and the inward gradient is clearly visible and points to the upper left corner.

Our classes and the \ion{H}{II} regions identified by \citet{James2020} agree well  (Fig.~\ref{Fig:JKB18classmap_sup}). Because James et al. only used the H$_\alpha$ emission line , they were unable to distinguish planetary nebulae, as we do in our density-excitation diagram above (Fig.~\ref{Fig:JKB18Polcaro}). Interestingly, most of their regions encompass at least two of our classes (classes 4 and 6), and their main structures contain the four classes (4, 6, 7, and 8) we found as \ion{H}{II} regions from the electron density-excitation diagram in Fig.\ref{Fig:JKB18Polcaro}. 

We find a heterogeneity (Fig.~\ref{Fig:JKB18O3N2}) in the metallicity Z(O$_3$N$_2$) indicator \citep[equation 1 of ][]{James2020}. This heterogeneity differs from that found by \citet{James2020}: 
It is between and within the \HII\ regions and the interstellar medium, and not in the large \HII\ regions along the proposed spiral arms.

There is a double dichotomy in this galaxy on each side of a line approximately from SE to NW that passes the foreground star: In the SW quadrant, the velocities are lower and the \ion{H}{II} regions are not or only weakly surrounded by gas. This dichotomy was noted by \citet{James2020}. We find that our class 1 is mainly in this quadrant, while class 3 is more present in the NE part. The differences between the median spectra of these two classes are weak. These two classes are also found all around the field of view and could be due to residuals of sky lines.

The spot of our class 9 is visible in the continuous map in Figure 5 of \citet{James2020}, but is not really visible in any other of their maps of emission lines or line ratios. According to its position in our electron density-excitation diagram (Fig.~\ref{Fig:JKB18Polcaro}), it is compatible with an \ion{H}{II} region or a planetary nebula. The continuum is much redder than that of the other \ion{H}{II} region classes 4, 6, 7, and 8, however, suggesting that the spectrum is dominated by dense gas. Furthermore, the spot possess a unique emission line around 4873-4874 \AA\ that could be a \ion{Fe}{II} or a \ion{N}{III} line.
Since the spectra of JKB~18 were not normalised in our study, the clustering analysis is sensitive to the global level of the continuum, that is, to the mass of the emitting source. The continuum in class 9 is higher than for the other classes (Fig.~\ref{Fig:JKB18_AllMedSpec}), implying that the spot is a denser region. 

The choice not to normalise the spectra is justified in the case of JKB~18 because we do not expect much variability in the masses or densities on the scale of the spaxels. This does not hold for more complex objects such as NCG1068 and NGC~4151, which are presented below, and it clearly does not hold in the case of integrated spectra of galaxies \citep{Fraix-Burnet2021,Dubois2022}.

\section{Analysis of NGC~1068}
\label{NGC1068}

\subsection{Results}
\label{NGC1068_results}

\subsubsection{Results for the whole data cube}
\label{NGC1068_Results_GeneralStudy}

\begin{figure}
        \includegraphics[width=\linewidth]{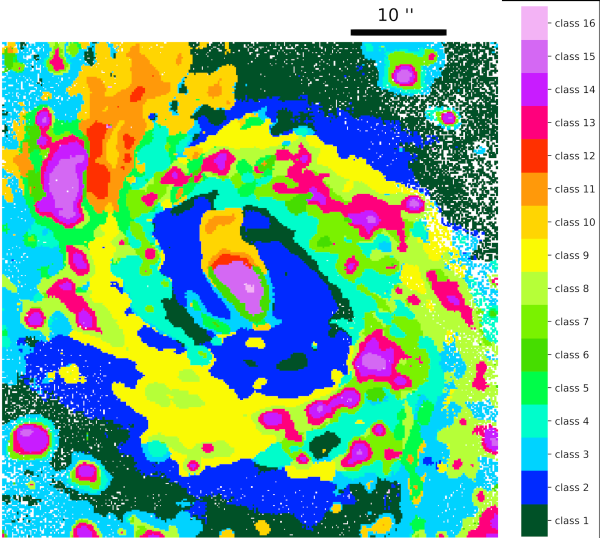}
        \caption{Class map for NGC~1068. The class nomenclature is explained in Fig.~\ref{Fig:JKB18classmap}.} 
        \label{Fig:NGC1068ClassmapK16}
\end{figure}

An optimum of 16 classes was found for NGC~1068. Many complex morphological structures are detected (Fig.~\ref{Fig:NGC1068ClassmapK16}):
an elongated central region, a ring, a SW string, a more complex NE structure, and several clumps outside the ring.
Every class extends over at least a few spaxels. 

The NE complex structure is made of two distinct regions: an extended region made of classes 10, 11, and 12, and a more compact region made of classes 13, 14 and 15. 

The NGC~1068 ring  hosts the most substantial diversity of classes: 4 to 9, and 13 to 15. The inner border of the ring is mostly composed of class 4 and some of class 9, whereas its outer border is composed of class 9 and some of class 4. 

The SW string and the clumps outside the ring are essentially composed of the same classes as the ring, but there is a noticeable difference in a diffuse component encompassing these two types of structures: class 3 approximately delineates the SW string and the clumps outside the ring, while class 2 is everywhere else, from the outer ring to the central region.

The central region is elongated towards the NE-SW axis, with a gradient of well-separated classes: 16, 15, 12, 11, and 10 towards the NE, and 16, 15, 6, 5, and 10 towards the SW. Another gradient lies in the perpendicular direction: 
16, 15, 6, and 3 along the NW-SE axis. 

The inner part of the central region is composed of two classes: 16 and 15. While class 16 is unique to the disc and confined to the very central core, class 15 is found in other parts of the galaxy, which is somewhat puzzling. In order to investigate further differences in the spectra of the spaxels within class 15, we subclassified this class. This is described in Sect.~\ref{NGC1068_Results_Nucleus} below.

The median spectra of the 16 classes (Fig.~\ref{Fig:NGC1068_AllMedSpec}) show a huge variability between the classes in the [\ion{O}{III}], H$_\alpha$, H$_\beta$, and [\ion{S}{II}], but present a low dispersion in their continuum. Classes 1 to 3 present a lower relative intensity notably in H$_\alpha$ and H$_\beta$. Classes 5, 6, and 10 to 12 have increasingly intense [\ion{O}{III}] emission lines. Classes 4, 7, 8, 9, 13, and 14 have high $H_\alpha$ but relatively low [\ion{O}{III}]. Class 15 presents intense H$_\alpha$ and $H_\beta$ and a huge variability in [\ion{O}{III}] both in intensity and width. Class 16 has intense and broad lines, but they are slightly shifted.

\begin{figure}
        \includegraphics[width=\linewidth]{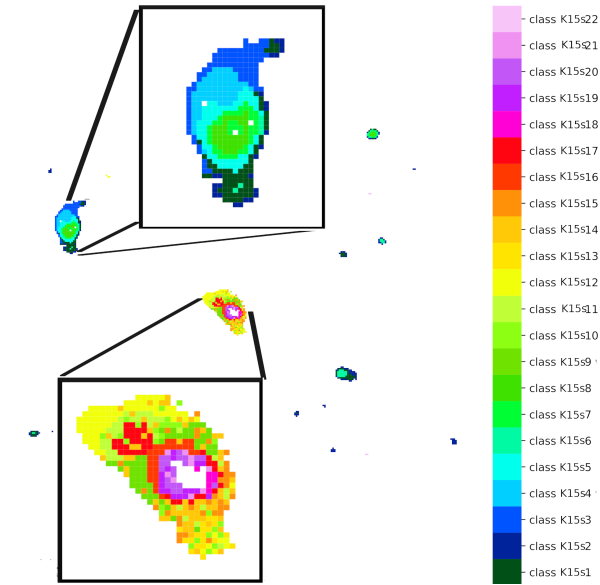}
        \caption{Class map for the NGC~1068 subclassification of class 15 with aligned spectra.}
        \label{Fig:ClassImageK22_SubclassificationK15_pretreated}
\end{figure}

\begin{figure}
        \includegraphics[width=\linewidth]{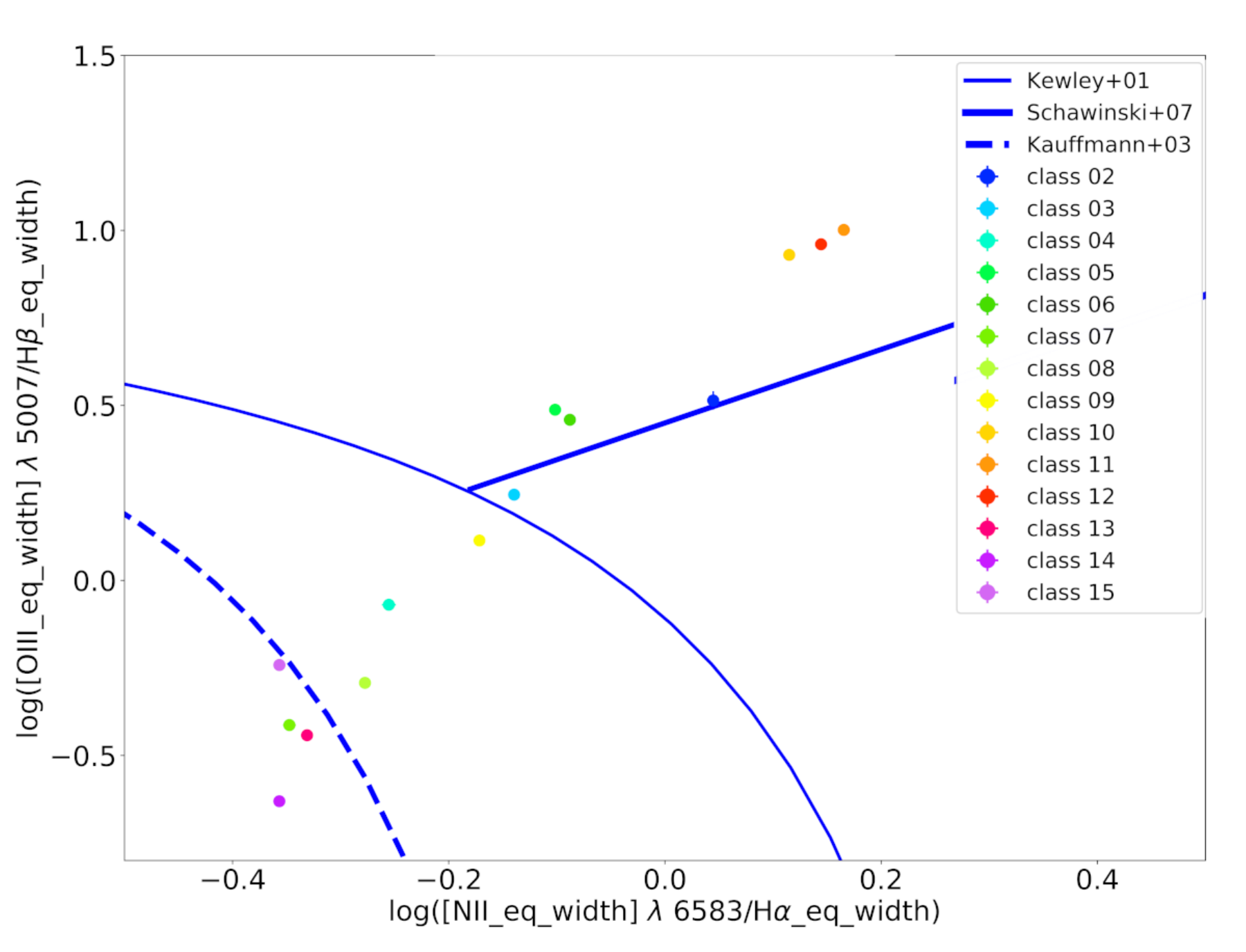}
        \caption{BPT diagram for NGC~1068. The class 1 median spectrum lacks one of the lines (H$_\beta$) and so does not appear in this plot. The error bars are the same as in Fig.~\ref{Fig:JKB18Polcaro}.}
        \label{Fig:NGC1068BptDiagK16}
\end{figure}

\begin{figure}
        \includegraphics[width=\linewidth]{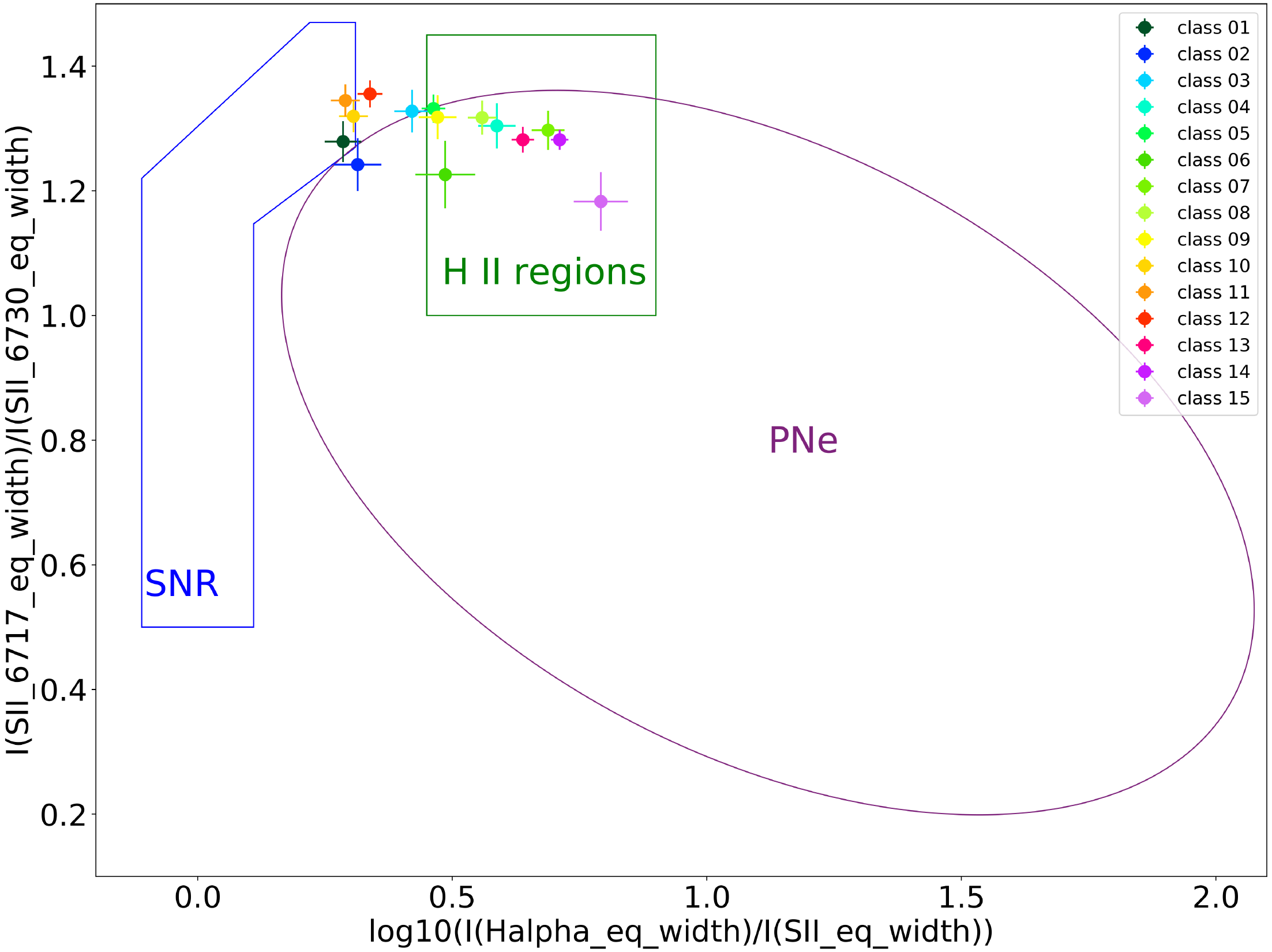}
        \caption{Electron density-excitation diagram for NGC~1068. The error bars are the same as in Fig.~\ref{Fig:JKB18Polcaro}.}
        \label{Fig:NGC1068PolcaroDiagK16}
\end{figure}

\subsubsection{Subclassification of class 15} 
\label{NGC1068_Results_Nucleus}

Class 15 is found in distinct parts of the galaxy, and its spectra show some peculiar behaviour for the [\ion{O}{III}] lines, that is, a high dispersion and a broad base with a thinner line on top of it. This suggests that this class could gather distinct physical regions. We therefore subclassified this class to investigate the more subtle differences in the spectra (Sect.~\ref{method:subclassification}) This led to an optimum of 22 subclasses (Fig.~\ref{Fig:ClassImageK22_SubclassificationK15_pretreated}). 
For clarity, we call these subclasses K15s\textit{XX}, where \textit{XX} extends from 1 to 22.
The subclasses clearly separate the nucleus (subclasses K15s9 to K15s21) from other regions (subclasses K15s1 to K15s8). In the latter, a gradient of the subclasses is visible. Subclass K15s22 is made of very few spaxels with very noisy spectra.
 Subclasses K15s3 and K15s4 are only found in the NE structure. 
 
Into the nucleus, the subclasses form two main adjacent roundish structures: subclasses K15s17 and K15s11 on the NE, and subclasses K15s16, K15s18, K15s19, K15s20, and K15s21 at the centre.  The extreme NE of the nucleus homogeneously consists of the subclass K15s12. The SW of the nucleus is more noisy, with subclasses K15s15, K15s14, and K15s13.

The median spectra of the nucleus subclasses (Fig.~\ref{Fig:NGC1068_AllMedSpecSubclassif}) show a dichotomy in line widths and [\ion{O}{III}] intensity: subclasses K15s1 to K15s8 have narrow lines and low [\ion{O}{III}], while subclasses K15s9 to Ks21 have broad lines and intense [\ion{O}{III}]. The median spectra of subclass K15s22 are noisy. K15s1 to K15s8 have intense H$_\alpha$, H$_\beta$, [\ion{N}{II}], [\ion{S}{II}] and [\ion{S}{III}], but they present variability along these lines. Notably, K15s3 and K15s4 have a shallower [\ion{S}{III}] line. 
The median spectra of subclasses K15s9 to K15s21 are variable in the [\ion{S}{II}] emission with an overall increase for the [\ion{S}{II}] at 6730~\AA,\ but decrease from K15s9 to K15s13, followed by an increase from K15s13 to K15s16 before a decrease from K15s18 to K15s21. The latter behaviour is seen in H$_\alpha$, H$_\beta$ and can be extended to [\ion{N}{II}] to some degree.

Subclasses K15s16 and K15s13 show surprising spectra. Some emission lines have two peaks, the second of which is blueshifted: [\ion{S}{III}], [\ion{Ar}{III}], [\ion{O}{III}] 4958 \AA\, and H$_\beta$. While being fainter, the same behaviour is detected in subclass K15s20. This might have misled the automatic computing of the redshift, so that we subclassified the same normalised spectra of class 15, but did not correct them for the kinematics. The result in Fig.~\ref{Fig:class_image_K18_NoDeredshifted} is fully consistent with the result with aligned spectra (Fig.~\ref{Fig:ClassImageK22_SubclassificationK15_pretreated}). The classmap appears to be less noisy in the central blob, however, which might be due to the influence of the kinematics (see Sect.~\ref{Ap:NGC1068undereshifted} and Fig.~\ref{Fig:class_image_K20_NoDeredshifted}).

\subsection{Discussion}
\label{NGC1068_discussion}

\subsubsection{Ionisation sources of the classes}

The BPT diagram (Fig.~\ref{Fig:NGC1068BptDiagK16}) shows that classes 7, 13, 14, and 15 are of the \HII\ type, classes 4, 8, and 9 are composite, and classes 5, 6, 10, 11, and 12 are AGNs. Classes 2 and 3 are at the limit between low-ionisation nuclear emission-line region (LINERs) and AGNs. Class 1 has no H$_\beta$ line detected by ALFA and thus cannot be placed in this diagram. Class 16, situated at the very centre of the galaxy, is absent as well because it has very wide lines, so that their automatic identification is very difficult. However, the median spectrum of this class is undoubtedly that of an AGN.

The electron density-excitation diagram (Fig.~\ref{Fig:NGC1068PolcaroDiagK16}) confirms the \HII\ classes (7, 13, 14, and 15) found in the BPT diagram, but shows variability in the H$_\alpha/[\ion{S}{II}$] line ratio. The composite classes (4, 8, and 9) and the two AGN classes 5 and 6 now appear as \HII. The other AGN classes (10, 11, and 12) seem closer to being SNR, that is, shock ionised. Classes 1 and 2 are SNR and planetary nebulae, respectively, but the electron density-excitation diagram may lead to a misinterpretation for extended regions that are observed at high resolution, such as in the case of spaxels \citep[e.g. ][]{Akras2020}.

\subsubsection{Ring}

\begin{figure}
        \includegraphics[width=\linewidth]{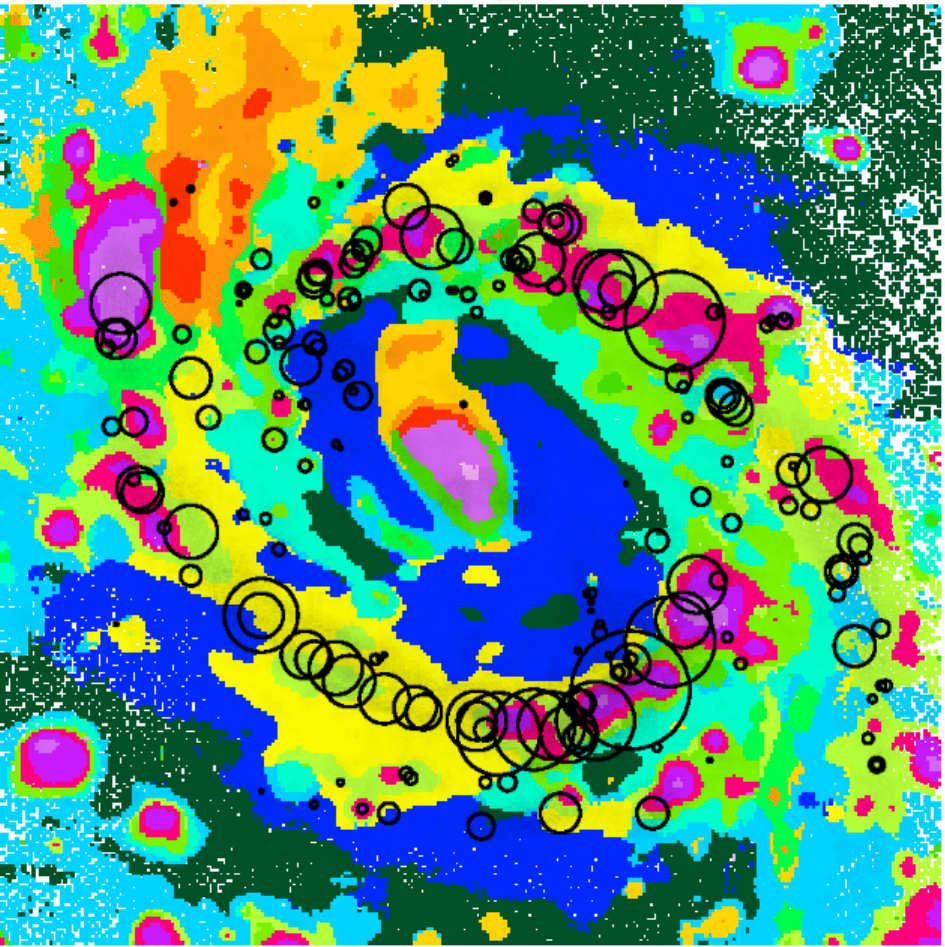}
        \caption{Class map for NGC~1068 superposed with giant molecular clouds identified by \citet{10.1093/pasj/psw122}. }
        \label{Fig:NGC1068ClassmapK16_sup_GMC}
\end{figure}

\begin{figure}
        \includegraphics[width=\linewidth]{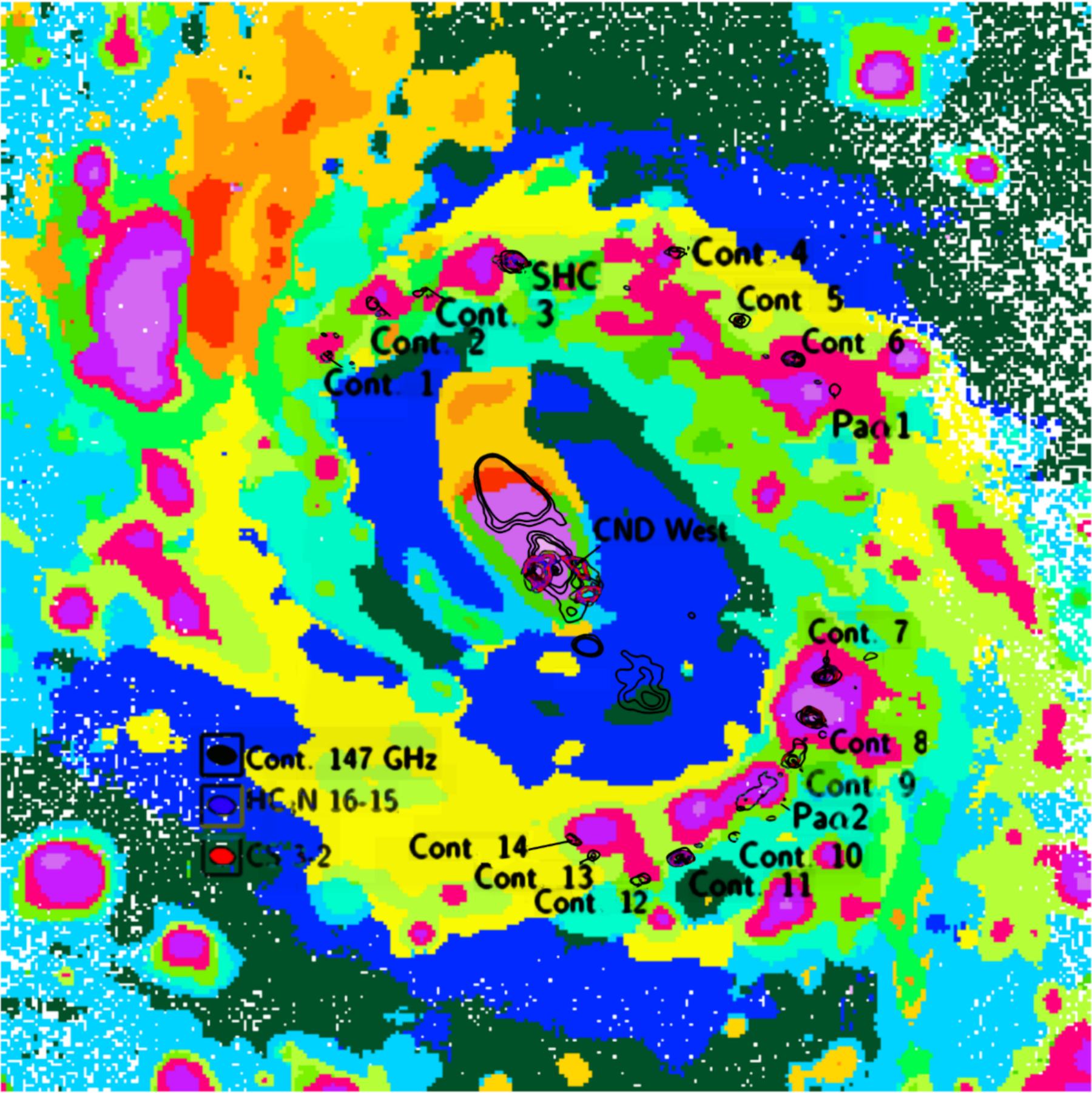}
        \caption{Class map for NGC~1068 superposed with HC$_3$N emission, CS 3-2 emission, and the 147~Ghz continuum found by \citet{10.1093/mnras/stab197}.}
        \label{Fig:ClassImageK16_superposed_H3CNRico2021}
\end{figure}

Our \HII\ classes 13 to 15 in the core of the ring are thus star-forming regions. They match the giant molecular clouds as traced by \trCO\  \citep[Fig.~\ref{Fig:NGC1068ClassmapK16_sup_GMC}; ][]{10.1093/pasj/psw122} quite well in general. The noticeable exception is the SE region of the ring, where many molecular clouds lie in our extended classes 8 and 9. These regions might be too opaque to visible light, or star formation might not be active enough. The agreement with the HC$_3$N and CS 3-2 emission map \citep[Fig.~\ref{Fig:ClassImageK16_superposed_H3CNRico2021}; ][]{10.1093/mnras/stab197} is also quite good. The extension of our classes is larger, and there are some mismatches in the central localisation. 

A perfect match between single-line observations and multivariate classification at different wavelength domains cannot be expected. Our analysis correctly reveals the core of the ring and its associated star formation regions, however. We also find gradients of classes where emission lines characteristic of \ion{H}{II} regions are enhanced.

\begin{figure}
        \includegraphics[width=\linewidth]{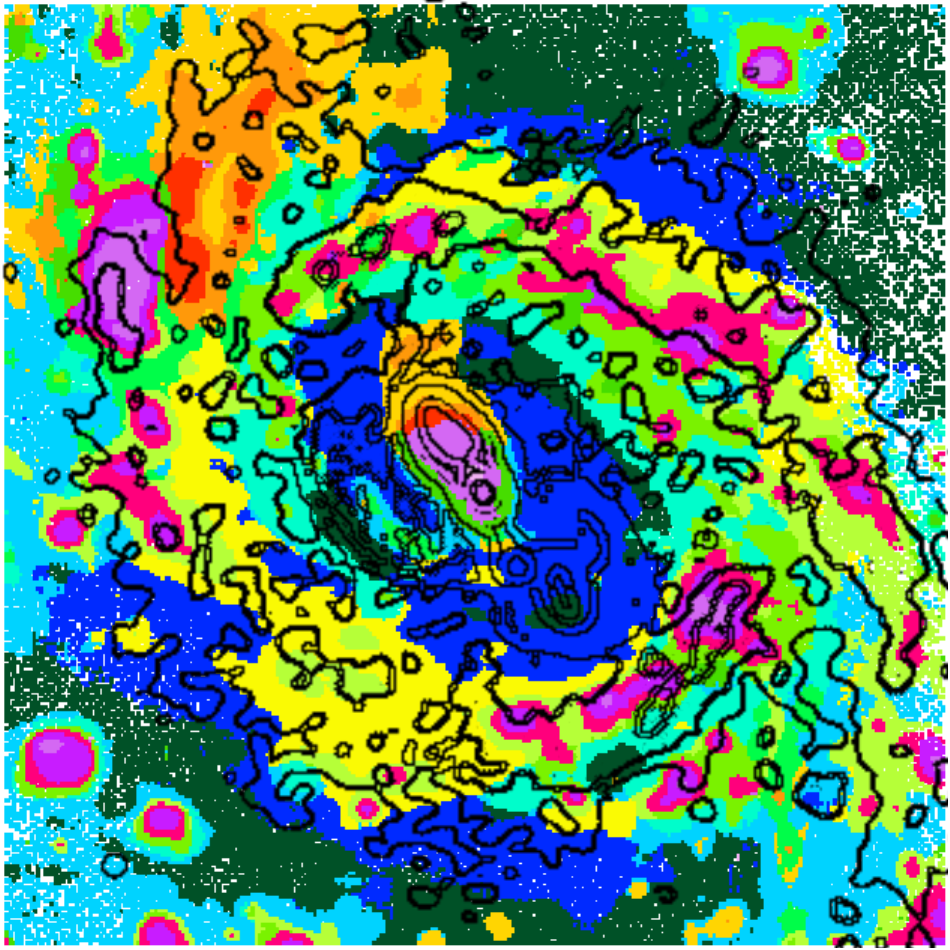}
        \caption{Class map for NGC~1068 superposed with the VLA contours at 18~cm from \citet{Gallimore1996}.}
        \label{Fig:ClassImageK16_superposed_Gallimore1996}
\end{figure}

\begin{figure}
        \includegraphics[width=\linewidth]{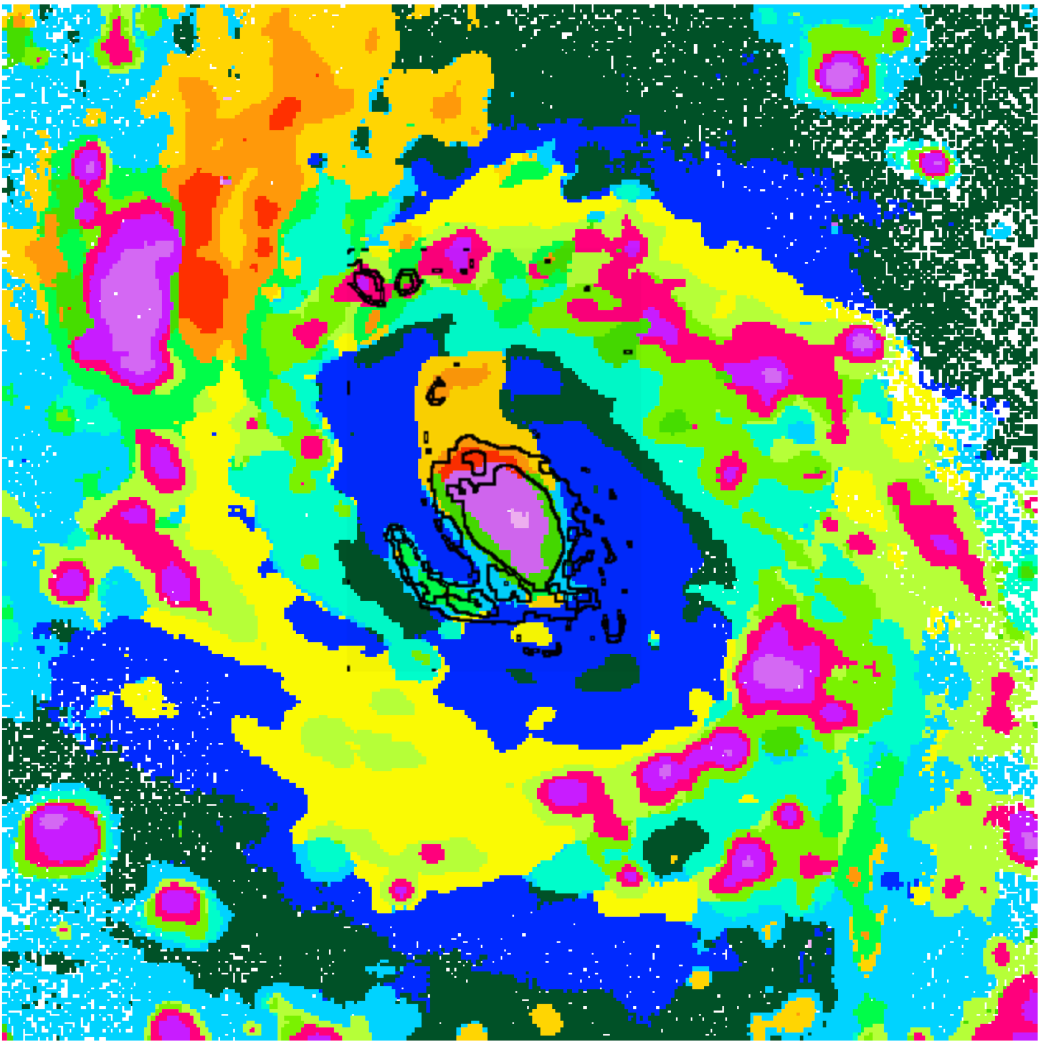}
        \caption{Class map for NGC~1068 superposed with the contours of the $H_\alpha + [\ion{N}{II}]$ map by \citet{Capetti_1997}.}\label{Fig:ClassImageK16_superposed_HalphaNIICapetti1996}
\end{figure}

\begin{figure}
        \includegraphics[width=\linewidth]{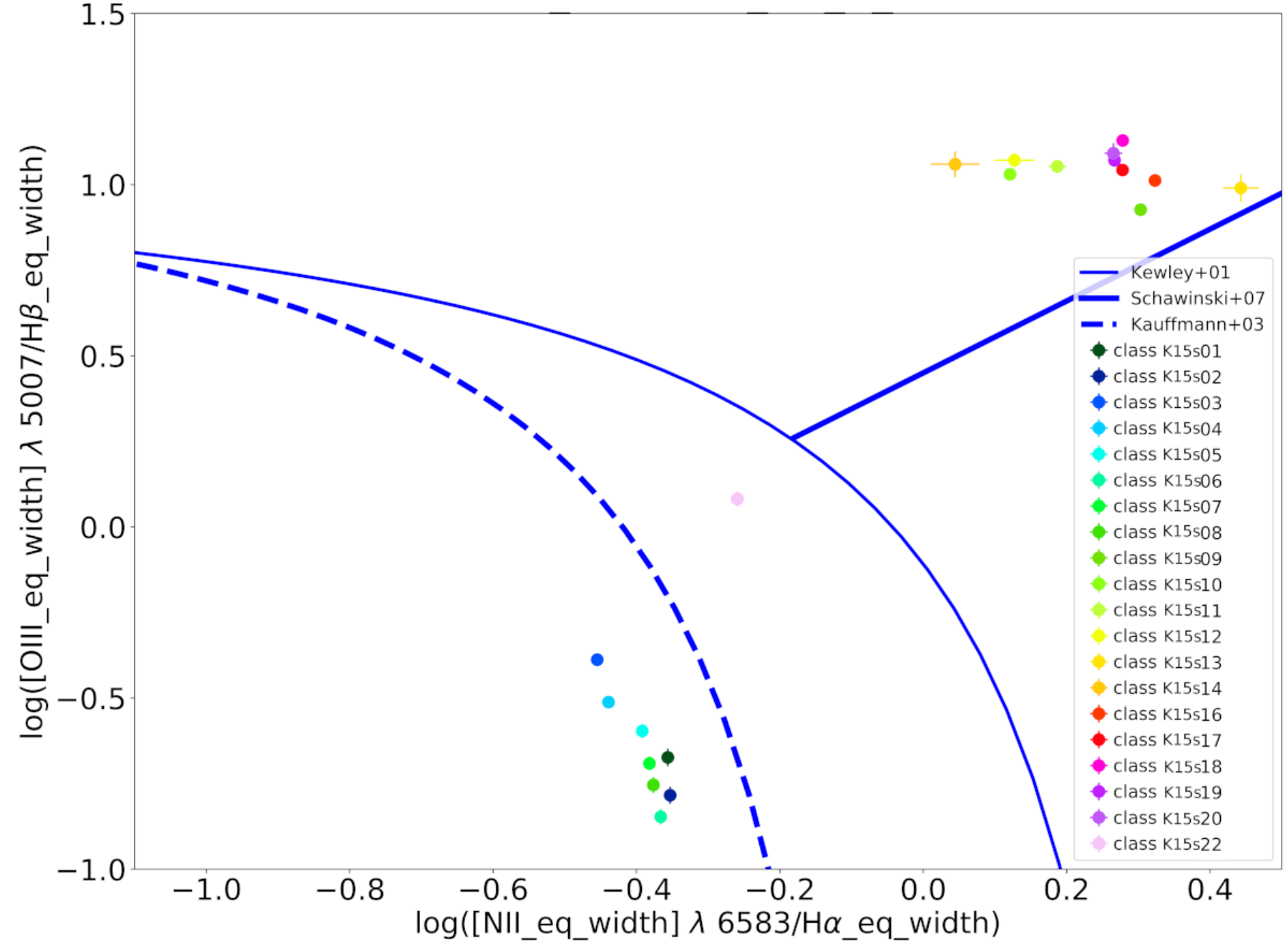}
        \caption{BPT diagram of NGC~1068 subclassification of class 15 with aligned spectra. Subclasses K15s15 and K15s21 are absent because ALFA had difficulties deblending the [\ion{N}{II}] at 6583~\AA\, but they are safely classified as AGN in OI BPT.  The error bars are the same as in Fig.~\ref{Fig:JKB18Polcaro}.}
        \label{Fig:BptDiagK22_SubclassificationK15_pretreated}
\end{figure}

\subsubsection{Elongated structures}

Our AGN classes 10, 11, and 12 are clearly aligned or elongated towards the NE, both in the central part and in the extended NE structure outside the ring. This incidentally is the direction of the well-known radio SW-NE jet, so that these classes may be related to the biconical cones due to the interaction of the jet with the insterstellar medium (ISM) where AGN ionisation dominates \citep{Gallimore1996}. This also agrees very well with the [\OIII]\ maps from \citet{Venturi2021}.

In the central part, the different gradient structures 10, 11, and 12 observed NE and 6, 5, and 10 SW of the nucleus are superposed with the observed jet in the radio by \citet{Gallimore1996} (Fig.~\ref{Fig:ClassImageK16_superposed_Gallimore1996}). In addition, class 6, which borders the nucleus except to the N, corresponds to the last contour of the jet.

The H$_\alpha$/[\ion{S}{II}] ratio, indicating the amount of ionisation due to shocks, increases perpendicular to the jet axis in the gradient 15, 6, and 3 towards the NW and SE and increases along the jet axis in the gradient 15, 6, 4, and 10 towards the SW (Fig.~\ref{Fig:NGC1068PolcaroDiagK16}). Towards the NE, this ratio is very similar for classes 10, 11, and 12. This places the first two classes within the supernovae remnant zone of the electron density-excitation diagram, class 12 being just outside. This appears to agree with shock signatures found by \citet{Venturi2021}.

Out of the five AGN classes in the BPT diagram, two groups emerge both spatially and physically: classes 10, 11, 
and 12 show a higher ionisation degree and form larger structures along the jet axis than classes 5 and 6. Likewise, in the gradients in \HII\ regions, there is an AGN-dominated gradient of classes 10, 11, and 12 corresponding to an increase of all the emission lines: inside the ring, the gradient 10, 11, and 12 is orientated towards the AGN nucleus along the jet axis, whereas outside the ring in the extended region to the NE, the gradient 10, 11, and 12 follows the axis perpendicular to the jet towards the SE.

The structures that we find in the central region are similar to the H$_\alpha$ + [\ion{N}{II}] map \citep[Fig.~\ref{Fig:ClassImageK16_superposed_HalphaNIICapetti1996}, ][]{Capetti_1997}. In particular, the arc-like structure to the SE that is composed of classes 3, 5, and 6 matches the structure seen in the H$_\alpha$ + [\ion{N}{II}] image perfectly.
There are two gradients of the classes, 6-3 in one direction, and 15, 12, and 11 in the other (Fig.~\ref{Fig:NGC1068ClassmapK16}), which correspond to a decrease in H$_\alpha$ + [\ion{N}{II}].  Classes 15 and 12 are within the radio lobe edge in the image of \citet{Capetti_1997}, and class 10 corresponds to their high-ionisation cap (Fig.~\ref{Fig:ClassImageK16_superposed_HalphaNIICapetti1996}).

Class 15 is present in the inner central part as well as in the compact NE region outside the ring. These two regions are spatially very distinct, so that they are probably physically different as well. The median spectrum for this class lacks the strong or broad emission lines typical of the nucleus ([\ion{O}{III}], H$_\alpha$, [\ion{N}{II}], [\ion{S}{II}], and [\ion{S}{III}]), but the dispersion of the spectra within the class in these emission lines is rather large. The subclassification of this class (Fig.~\ref{Fig:ClassImageK22_SubclassificationK15_pretreated}) resulted in a clear separation between the two regions, with AGN classes in the nucleus and the \HII\ classes in the outer NE compact region (Fig.~\ref{Fig:BptDiagK22_SubclassificationK15_pretreated}). They are discussed in the following two sections.

\subsubsection{Compact NE region outside the ring}

Outside the nucleus, only classes K15s1 to K15s8 are present, and are all found to be \HII\ regions (Fig:~\ref{Fig:BptDiagK22_SubclassificationK15_pretreated}). Most of them are composed of classes K15s1 and K15s2, with a central region in which K15s6 forms a clear gradient in the BPT diagram. Classes K15s3 and K15s4 are specific to the compact NE structure, however. These two classes have a higher [\OIII]/H$_\beta$ ratio than the other \HII\ classes. 

This compact NE region is thus an intense star-forming region and more extended than all the \HII\ regions that our classification finds in NGC~1068. It shows \trCO\ emission (Fig.~\ref{Fig:NGC1068ClassmapK16_sup_GMC}). It is apparently located in a spiral arm departing from the eastern side of the ring, but it is strikingly situated next to the extended NE region, which is possibly ionised by shocks due to the jet. It is also surrounded by the two AGN classes 5 and 6 (Fig.~\ref{Fig:NGC1068ClassmapK16}). This tends to suggest some connection with the presence of the jet (e.g. it might be the relic of the interaction of the ISM with the jet, spatially shifted after galaxy rotation), but it is difficult to conclude without a much more detailed analysis. This is beyond the scope of this paper.

\subsubsection{Nucleus}
\label{NGC1068_Discussion_NucleusStudy}

The subclassification reveals that according to the BPT diagnostic diagram (Fig.~\ref{Fig:BptDiagK22_SubclassificationK15_pretreated}),  all classes K15s9 to K15s21 of the nucleus are of AGN type.

The two roundish structures can be related to the four different gas blobs found by \citet{Shin_2021} from kinematics properties: classes K15s17 and K15s11 (to the NE) correspond to the R1 redshifted gas blob, and classes K15s16, K15s18, K15s19, K15s20, and K15s21 to the B1 blueshifted blob, which is at the centre of the galaxy. Moreover, the K15s17 structure we find coincides with a knot in the radio jet (Fig.~\ref{Fig:ClassImageK16_superposed_Gallimore1996}).
Class K15s16 lies at the edge of B1, and its spectra show double peaks in high-ionisation emission lines ([\ion{O}{III}], [\ion{Ar}{III}], and [\ion{S}{III}]), as do the spectra of class K15s13, which lies SW of B1.
The intriguing kinematics of the gas blobs caused \citet{Shin_2021} to propose an additional smaller AGN nucleus located between R1 and B1, that is, between our classes K15s17 and K15s16. We find a subtle asymmetry in our subclassification class map (Fig.~\ref{Fig:BptDiagK22_SubclassificationK15_pretreated}): the two classes K15s16 and K15s20 around the centre B1 blob are slightly shifted towards the NE, that is, towards the hypothetical second AGN, while classes K15s18, K15s14, and K15s13 are principally found SW of blob B1.

The SW part of blob B1 appears to be somewhat pixellised on our classmap (Fig.~\ref{Fig:BptDiagK22_SubclassificationK15_pretreated}). This can be explained by real differences or by an imperfect alignment of the spectra because of the complex emission line structures. To better understand the role of the kinematics in our classification, we subclassified class 15 without alignment. The result is shown in Appendix~\ref{Ap:NGC1068undereshifted} and confirms our findings above. In particular, the smoother appearance of the classes around the nucleus might at least partially be attributed to the difficulty of automatically computing the redshift when the kinematics creates complex line structures.

\section{Analysis of NGC~4151}
\label{NGC4151}

\begin{figure}
        \includegraphics[width=\linewidth]{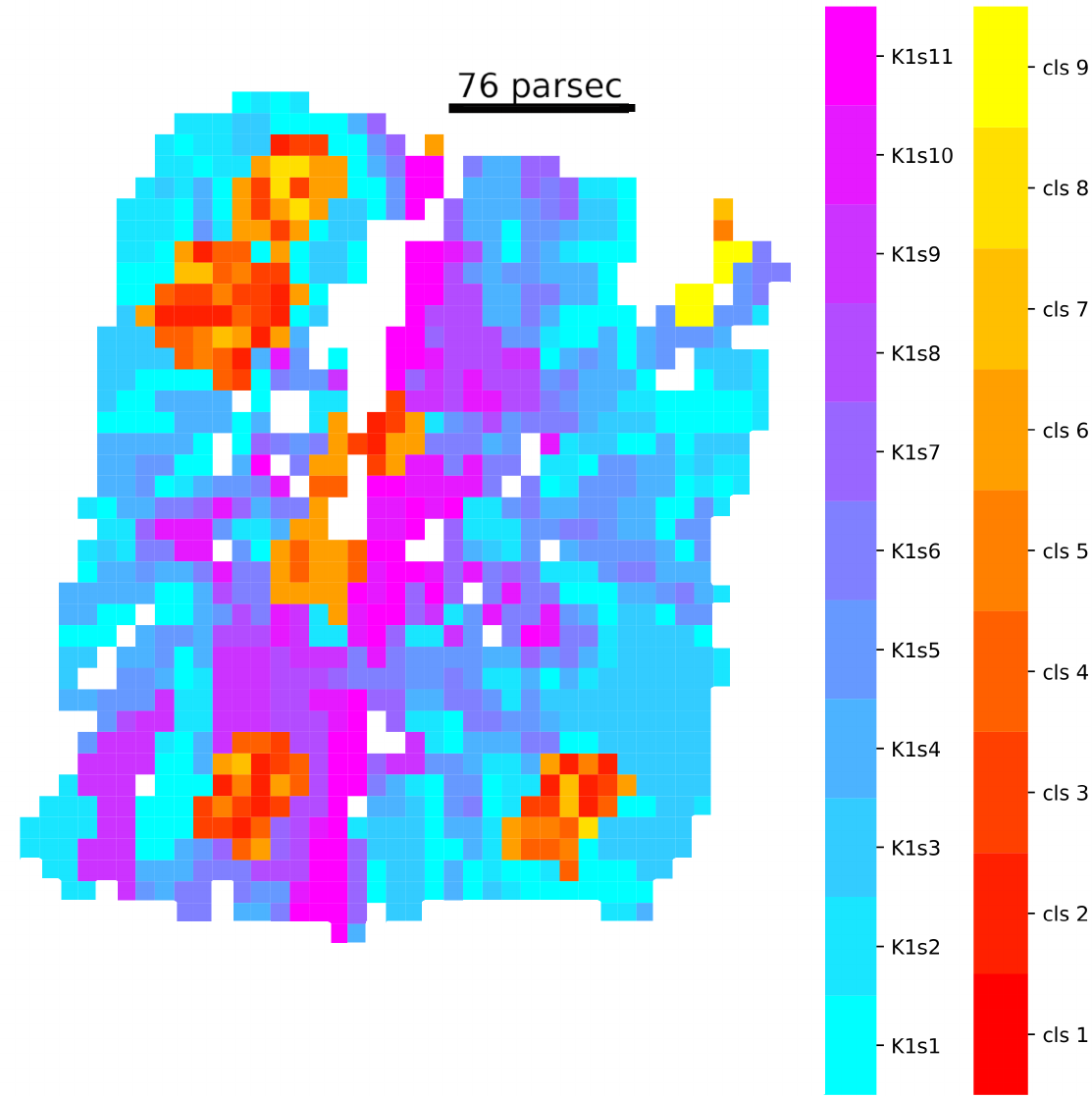}
        \caption{Class map summary for the Fisher-EM analysis of NGC~4151 with the classification into nine classes (reddish colours) and the subclassification of class 1 into 11 classes (blueish colours). The red class 1 is replaced by its subclassification and hence does not appear on this map. The white spaxels were removed because the automatic alignment failed. The class nomenclature is explained in Fig.~\ref{Fig:JKB18classmap}.}
        \label{Fig:NGC4151ClassmapSummaryFEM}
\end{figure}

\begin{figure*}
        \includegraphics[width=\linewidth]{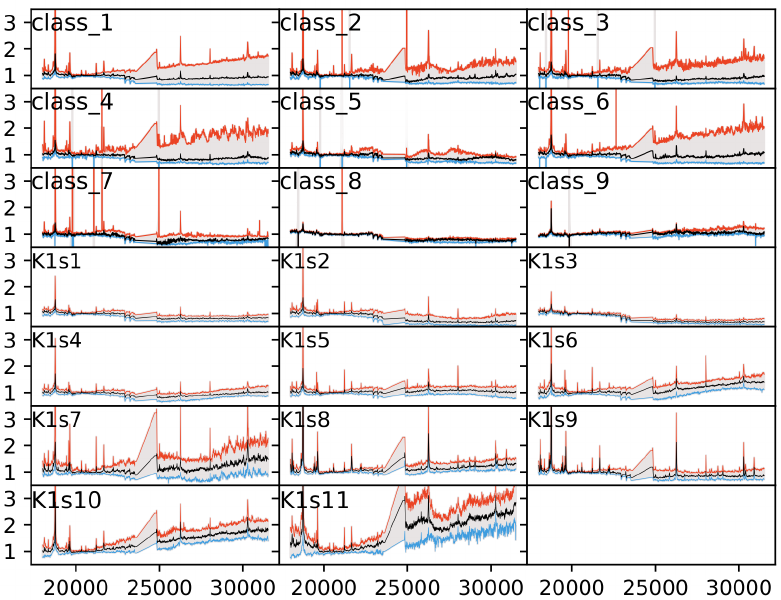}
        \caption{Median spectra (black) of every class and subclass for NGC~4151 with their dispersion (10 and 90\% quantiles in blue and red).}
        \label{Fig:NGC4151Allspectra}
\end{figure*}

\begin{figure}
        \includegraphics[width=\linewidth]{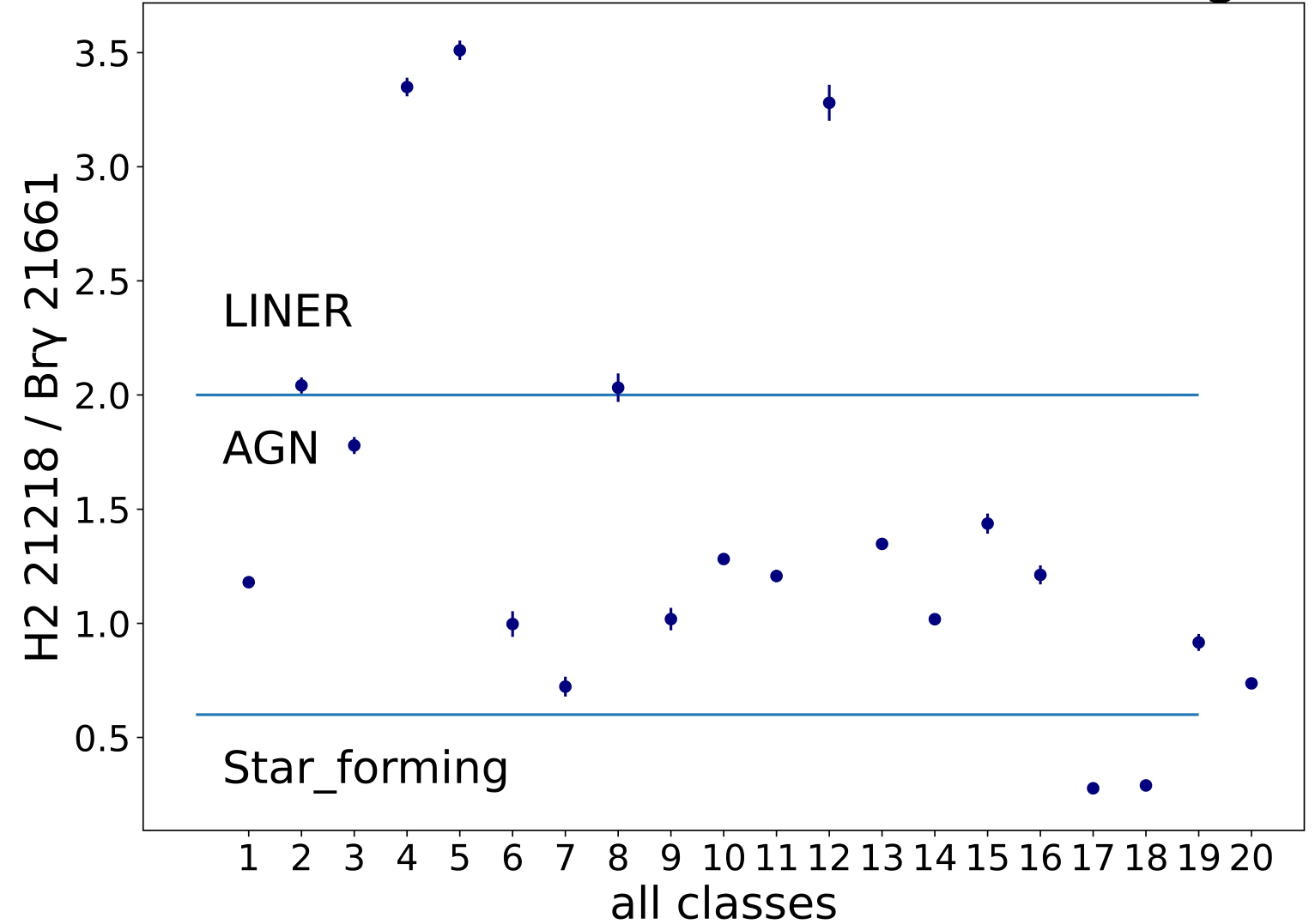}
        \caption{H$_2$ 21218 / Br$_\gamma$ ratio. Classes 1 to 9 correspond to the first classification with Fisher-EM, and classes 10 to 20 correspond to classes K1s1 to K1s11 of the subclassification step}
        \label{Fig:NGC4151H2Brgammaratio}
\end{figure}

\begin{figure}
        \includegraphics[width=\linewidth]{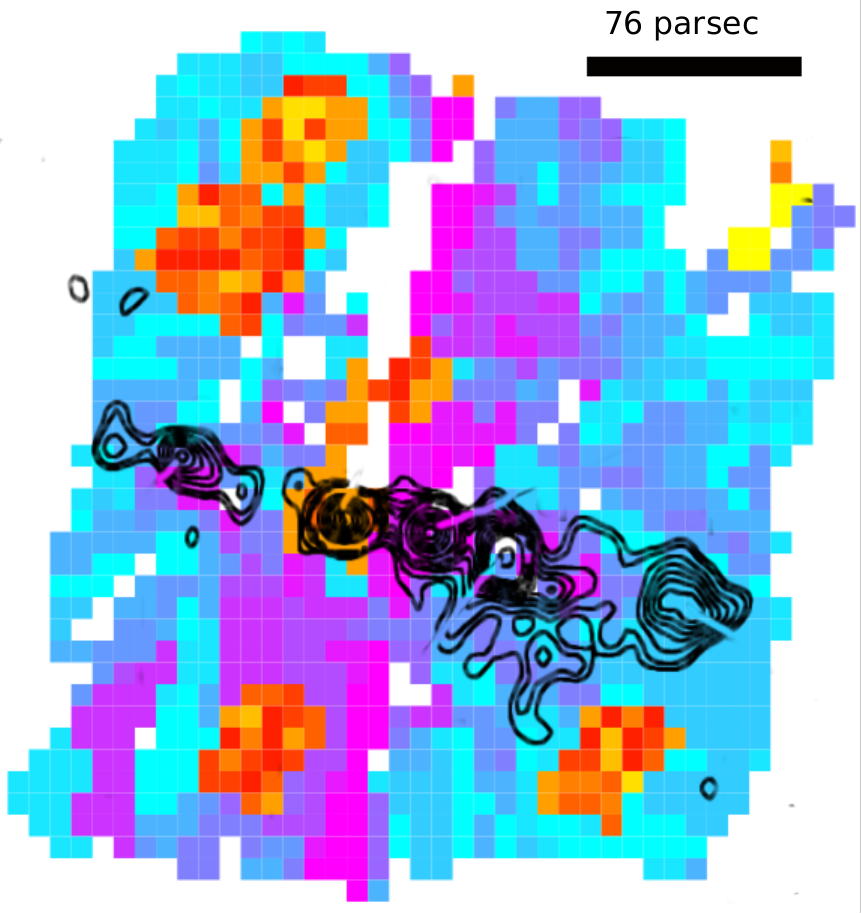}
        \caption{Class map for NGC~4151 with two subclassifications in Fisher-EM superposed with neutral hydrogen absorption from \citet{1995MNRAS.272..355M}. The colours are the same as in Fig.~\ref{Fig:NGC4151ClassmapSummaryFEM}}.
        \label{Fig:NGC4151AllClassmapFEM_supHIasborption}
\end{figure}

\begin{figure}
        \includegraphics[width=\linewidth]{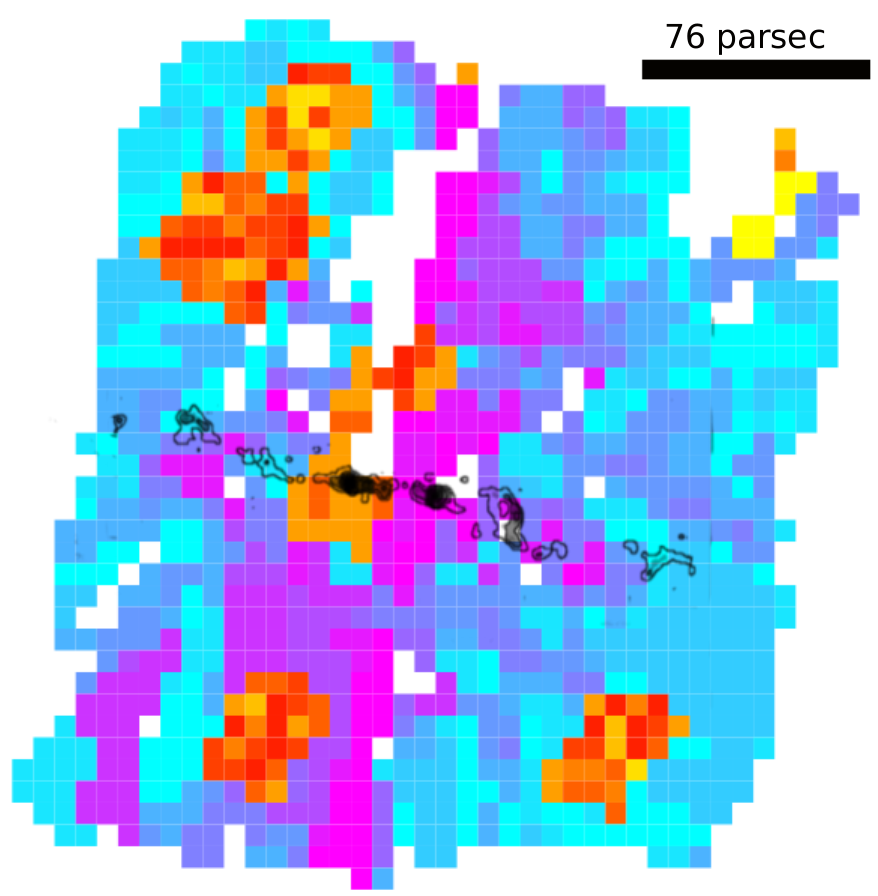}
        \caption{Class map for NGC~4151 with two subclassifications in Fisher-EM superposed with continuum at 21cm from \citet{2003ApJ...583..192M}. The colours are the same as in Fig.~\ref{Fig:NGC4151ClassmapSummaryFEM}}.
        \label{Fig:NGC4151AllClassmapFEM_sup21cm}
\end{figure}

\begin{figure}
        \includegraphics[width=\linewidth]{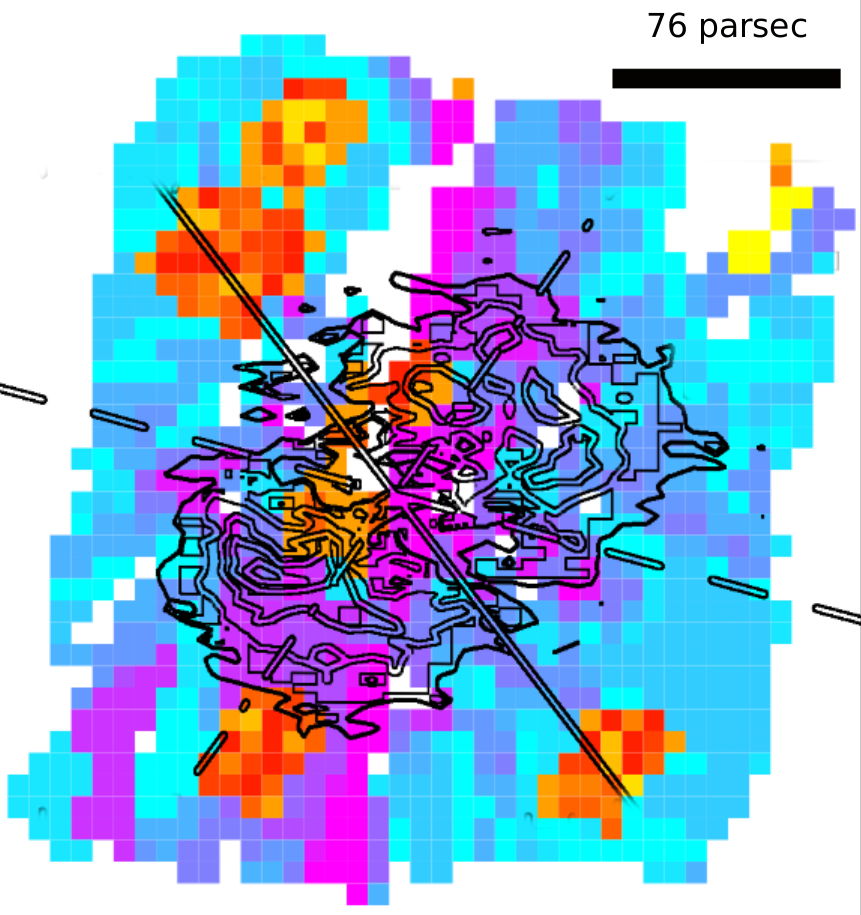}
        \caption{Class map for NGC~4151 with two subclassifications in Fisher-EM superposed with H2 emission from \citet{10.1111/j.1365-2966.2009.14388.x}. The continuous line shows the orientation of the major axis of the galaxy, the dashed line shows the orientation of the bicone, and the dot–dashed line shows the orientation of the bar. The colours are the same as in Fig.~\ref{Fig:NGC4151ClassmapSummaryFEM}}.
        \label{Fig:NGC4151AllClassmapFEM_supH2}
\end{figure}

\begin{figure}
        \includegraphics[width=\linewidth]{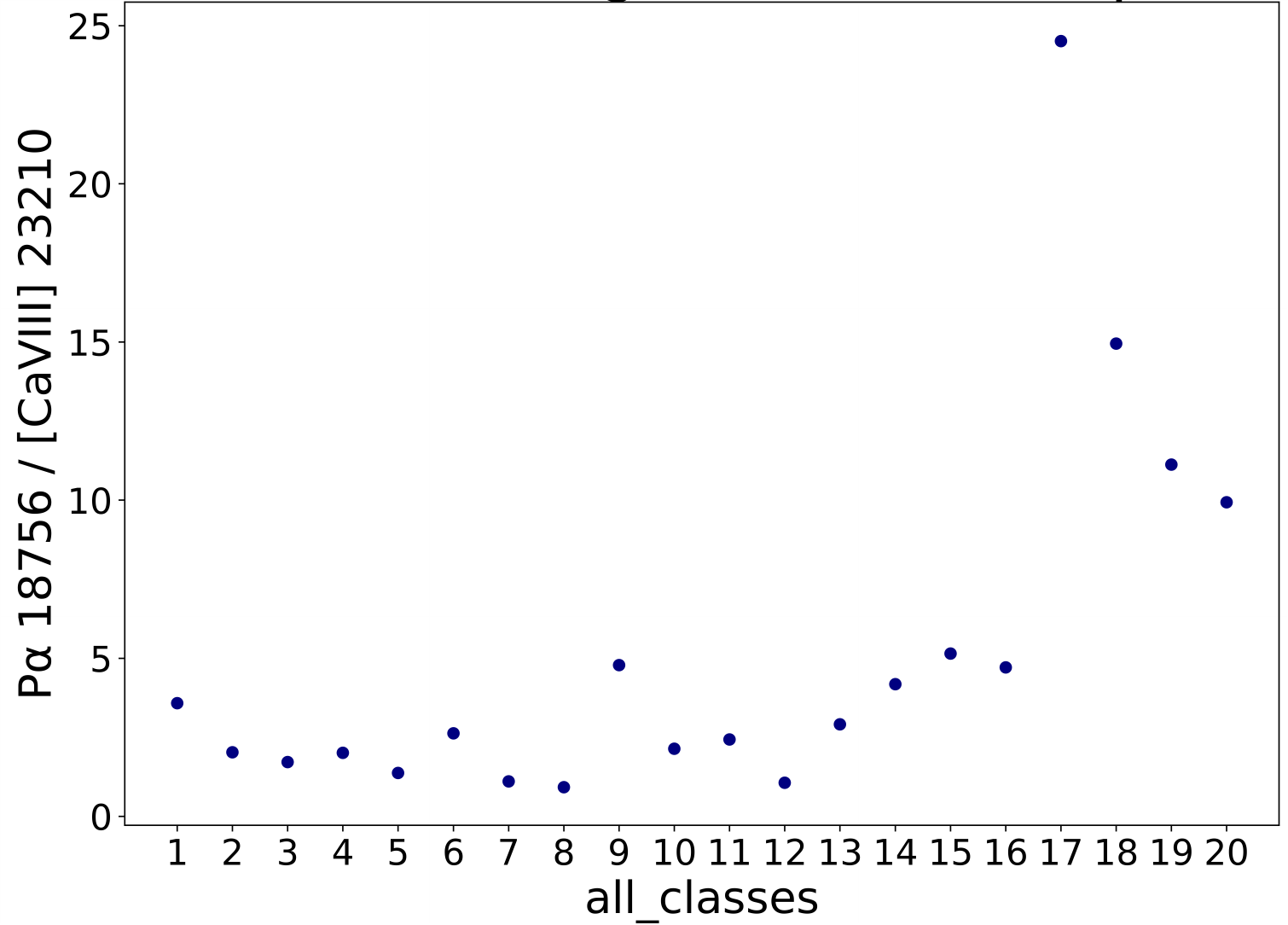}
        \caption{Pa$_\alpha$ 21218 / [\ion{Ca}{VIII}] ratio. Classes 1 to 9 correspond to the first classification with Fisher-EM, and classes 10 to 20 correspond to classes Ks1 to Ks11 of the subclassification step.}
        \label{Fig:NGC4151PaCaVII}
\end{figure}

\subsection{Results}
\label{NGC4151_results}

The classification of NGC~4151 with Fisher-EM led to an optimum of nine classes (Fig.~\ref{Fig:NGC4151ClassmapSummaryFEM}). Class 1 accounts for more than 90\% of the spaxels, and the eight other classes are found in five blobs: a central blob, a large blob to the NE, two blobs to the SE and SW, and a small blob to the NW.
The central blob is mainly composed of class 6, while the small NW blob is made of class 9. The composition of the other blobs is a composite of the other classes without coherent structures of more than five spaxels. 

The median spectra of the classes (Fig.~\ref{Fig:NGC4151Allspectra}) present variability in the Pa$_\alpha$, Br$_\beta$, and Br$_\gamma$ lines, the principal H$_2$ line (21218~\AA), the coronal lines [SiVI] and [MgVIII], and also in the continuum level and shape. On the one hand, class 1 to 4 and 6 present high dispersion both in the emission lines and in the continuum in the interval [24500,31500] \AA\ and a low dispersion in the interval [18000, 23500] \AA. On the other hand,  class 5 and 7 to 9 present a low dispersion in the continuum, but classes 5 and 7 present a high dispersion in their emission lines. 

Because of the high number of spaxels inside class 1 and of the significant dispersion of its spectra, we decided to subclassify this class. This led to an optimum of 11 subclasses forming coherent structures (Fig.~\ref{Fig:NGC4151ClassmapSummaryFEM}; for better visibility, Fig.~\ref{Fig:NGC4151_IndSubclassFem} shows maps of individual classes). In particular, there is a central linear structure (subclass K1s11) along the N-S axis and different blobs along the SE to NW axis composed of subclasses K1s2, K1s8, and K1s9. There is a noticeable individual elongated blob to the extreme SE made of subclass K1s9. A perpendicular structure, that is, E-W, seems to emerge with subclasses K1s6, K1s7, and K1s10. Finally, subclasses K1s1 to K1s5 seem to form a rough circular structure that is interrupted in the S. 

The median spectra of subclasses K1s1 to K1s5, K1s8, and K1s9 present a low dispersion in their continuum and a medium dispersion in their principal emission lines: the Pa$_\alpha$, Br$_\beta$, and Br$_\gamma$ lines, the principal H2 lines, and the coronal lines [SiVI] and [MgVIII]. Subclasses K1s6, K1s7, K1s10, and K1s11 show a greater dispersion both for the continuum and the principal emission lines. In addition, subclasses K1s7, K1s10, and K1s11 show a fainter or absent CO absorption band around 23000~\AA.

\subsection{Discussion}
\label{NGC4151_discussion}

The subclassification step for NGC~4151 is justified because one out of nine classes gathers 90\% of the spaxels. This proves to be clearly useful with the identification of several interesting structures. To confirm this result, we applied a slightly different algorithm, called HDDC, which attributes subspaces that are specific to each class. We expected this to result in a better distinction without a subclassification step. However, we find that the latter is also necessary. This algorithm and our analysis are presented in Sect.~\ref{App:HDDC}; the results are similar to those of Fisher-EM. 

The two classification steps made with Fisher-EM yield 20 classes altogether. To understand the nature of the different classes, we used the H$_2$/Br$_\gamma$ ratio (Fig.\ref{Fig:NGC4151H2Brgammaratio}). This ratio is commonly used in conjunction with the [Fe II] (12567 \AA), Pa$_\beta$ \citep[12818 \AA; ][]{Larkin_1998}, but we only used the threshold values of \citet{10.1111/j.1365-2966.2005.09638.x}. This ratio shows that 13 are classified as AGN (classes 1, 3, 6, 7, 9, K1s1, K1s2, K1s4 to K1s7, K1s10, and K1s11), 5 as LINERS (classes 2, 4, 5, 8, and K1s3), and 2 as star-forming (subclasses K1s8 and K1s9). The AGNs and LINERs are separated in the first classification, while the AGN and the star-forming classes are separated in the subclassification of class 1.

The H$_2$ 22247/ H$_2$ 21218 ratio shows that 19 of the 20 classes are compatible with the thermal excitation scenario for H$_2$, but classes 5, 8, and 9 are also compatible with the fluorescent excitation scenario, while class 7 fits no common standard \citep{1994ApJ...427..777M}. This result agrees with the various studies for NGC4151 \citep{10.1111/j.1365-2966.2009.14388.x}.

The central and NW blobs are mostly composed of AGN classes 6 and 9. The other blobs are a mix of AGN and LINER classes. The composite nature of the SW and NE blobs is consistent with the ionisation bi-cone axis. Surprisingly, the SE blob is also composite, but lies outside the bi-cones. However, an inflow of matter towards the centre could explain its ionisation properties \citep{May_2020}. The central blob is closest to the AGN and superposes a maximum in the neutral hydrogen absorption well \citep[Fig.\ref{Fig:NGC4151AllClassmapFEM_supHIasborption};][]{1995MNRAS.272..355M} or the 21~cm continuum map \citep[Fig.\ref{Fig:NGC4151AllClassmapFEM_sup21cm};][]{2003ApJ...583..192M}, confirming the pure AGN diagnostics above.

The subclassification of class 1 underlines the presence of star-forming regions along the NW-SE axis. They are located outside the centre. 
These regions agree with the H$_2$ emission \citep[Fig.~\ref{Fig:NGC4151AllClassmapFEM_supH2};][]{10.1111/j.1365-2966.2009.14388.x}. 

Class K1s3 is the largest LINER class. Its spatial distribution corresponds to the ionisation bi-cones. 

Class K1s11 forms a coherent elongated AGN structure along the N-S axis perpendicular to the jet direction. The axis of this structure is tilted counterclockwise with respect to the axis of the H$_2$ emission (Fig.~\ref{Fig:NGC4151AllClassmapFEM_supH2}). To our knowledge, a structure like this has not been reported before. Together with K1s10, K1s11 has the highest continuum in the rarely observed [24500-31500]\AA\ range. Apart from the star-forming classes (K1s8 and K1s9), they have the highest Pa$_\alpha$ emission (Fig.~\ref{Fig:NGC4151Allspectra}) and thus show a high Pa$_\alpha$/[CaVIII] ratio (Fig.~\ref{Fig:NGC4151PaCaVII}). In contrast to the other AGN classes, K1s10 and K1s11 lack a CO absorption band around 23000~\AA.

\section{Conclusion}
\label{Conclusion}

We have performed an unsupervised classification of spaxels for three galaxies using a GMM algorithm in a latent discriminative subspace called Fisher-EM. Our classes gathered similar spectra, and the mean or median spectrum of each class has a higher signal-to-noise ratio, which make it easier to interpret through spectral fitting or diagnostic diagrams.

In the three different types of galaxies we studied, we showed that the unsupervised classification of spaxels is not only feasible, but also useful to identify regions that have similar spectra based on all the information contained and not only on a few selected features. The interpretation of the classes is easily achieved from both their spatial localisation and some diagnostic diagrams, without the need to visualise many maps of different properties. In all cases, we identified new structures.

In JKB~18, we find 11 classes and identified many \HII\ regions that nearly all correspond to the previous description by \citet{James2020}. However, we were able to map these regions as extended, with gradients of ionisation intensities. We also identified more extended zones with much lower stellar formation and slightly different metallicities. In addition, we find a small new region that we call the spot, which might be a denser \HII\ region or a planetary nebula.

Our classification of spaxels in NGC~1068 yielded 16 classes. Using diagnostic diagrams, we find that some classes are of AGN type and some others are  \HII\ regions. Their spatial distribution corresponds perfectly to well-known structures such as spiral arms and a ring with giant molecular clouds. In particular, our \HII\ classes are preferentially located in the ring and the spiral arms, and they vary in the $H_\alpha/[\ion{S}{II}]$ line ratio. We also identified different classes in the inner and outer parts of the ring that the BPT diagram shows as composite regions, that is, photoionised by stars and/or by the central AGN.

We subclassified a class that contains the nucleus of NGC~1068 as well as extended structures far from it. We find 22 subclasses that are separated into two categories: the AGN classes lie around the nucleus, and the \HII\ regions lie in the extended structures. Two roundish structures and asymmetries are clearly visible inside the nucleus. Globally, our unsupervised classification of the NGC~1068 by the MUSE instrument helps to visualise the complex interaction of the AGN and the jet with the interstellar medium.

Finally, our analysis of the NIRSpec/JWST data cube for NGC~4151 yielded nine classes, and the subclassification of the class that gathered 90\% of the spaxels yielded 11 subclasses. Many structures can be identified and characterised by several emission lines and their continuum level and shape, mainly of AGN and LINER types. They can mostly be related to the jet interaction, the ionisation bicones, and the H$_2$ emission. We identified a new nearly linear structure, perpendicular to the jet and slightly tilted with the respect to the H$_2$ absorption map.

We have thus shown that unsupervised classification can be very useful to automatically identify spectroscopically identical regions in individual galaxies in IFS data cubes. The interpretation of the data only needs to be made on the mean or median spectrum of each class instead of on all spaxels. The map of the classes summarises many maps of different properties and highlights some structures with peculiar multivariate properties, such as gradients within \HII\ regions. Altogether, this exploratory work shows that the unsupervised classification of spaxels takes full advantage of the richness of information in the data cubes by presenting both the spectral and spatial information in a combined and synthetic way.

We have shown that the unsupervised classification of spectra requires them to be aligned (i.e. de-redshifted), otherwise the kinematics dominates the result. However, it is possible to visualise the internal motions within each class afterwards, that is, the detailed kinematics of spectroscopically identical regions.

We used some classical diagnostic diagrams to characterise our classes. These diagnostics are based on emission line ratios, while our classes are built from the entire spectrum including the continuum and the absorption or emission lines. We have shown in \citet{Fraix-Burnet2021} that the Fisher-EM classification of spectra is sensitive to line ratios as well. This means that the mean spectra obtained with an algorithm like this should be interpreted in their whole complexity, probably through model fitting. This was beyond the scope of this exploratory work.

Finally, we must mention that other maps obtained at different wavelengths can be added to complete the data cubes. This would require similar spatial samplings, but would advantageously include much more physics.

\begin{acknowledgements}
We thank an anonymous referee for their detailed and very constructive remarks that improved a lot the paper. We thank Roland Bacon and Pierre Ferruit for discussion prior to this project. 
The work on JKB~18 and NGC~1068 is based on data obtained from the ESO Science Archive Facility with DOI: https://doi.org/10.18727/archive/41. 
The work on NGC~4151 is based on observations made with the NASA/ESA/CSA James Webb Space Telescope. The data were obtained from the Mikulski Archive for Space Telescopes at the Space Telescope Science Institute, which is operated by the Association of Universities for Research in Astronomy, Inc., under NASA contract NAS 5-03127 for JWST. This observation is associated with program \#1364.
\end{acknowledgements}

\bibliographystyle{aa} 
\bibliography{MLIFS}

\begin{appendix}
\section{Median spectra of the classes for JKB~18}
    \label{Ap:JKB18}

\begin{figure*}
        \includegraphics[width=\linewidth]{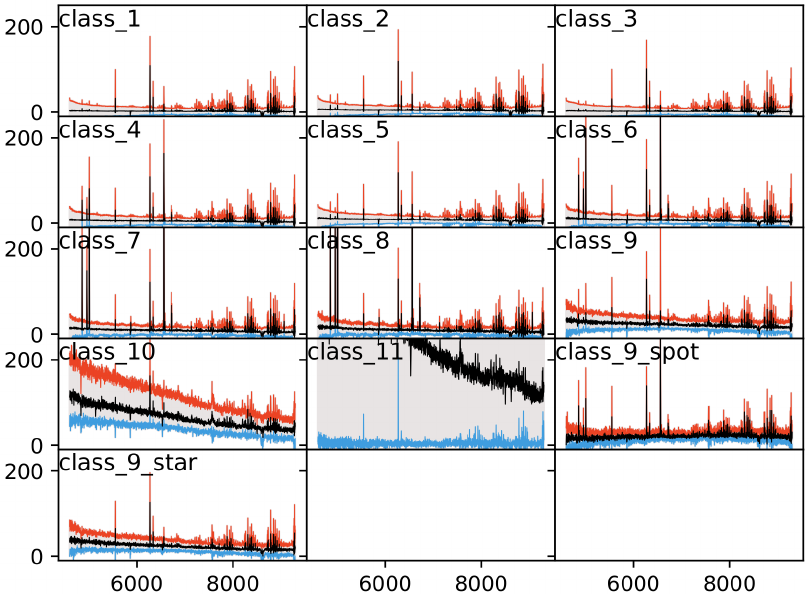}
        \caption{Median spectra (black) of every class for JKB18 with their dispersion (10 and 90\% quantiles in blue and red). The y-axis is in arbitrary units with the same scale for all plots.}
        \label{Fig:JKB18_AllMedSpec}
\end{figure*}

In this section, we show (Fig.~\ref{Fig:JKB18_AllMedSpec}) the spectra of the classes from the classification described in Sect.~\ref{JKB18_results}.

\section{Median spectra of the classes for NGC~1068}
    \label{Ap:NGC1068}

\begin{figure*}
        \includegraphics[width=\linewidth]{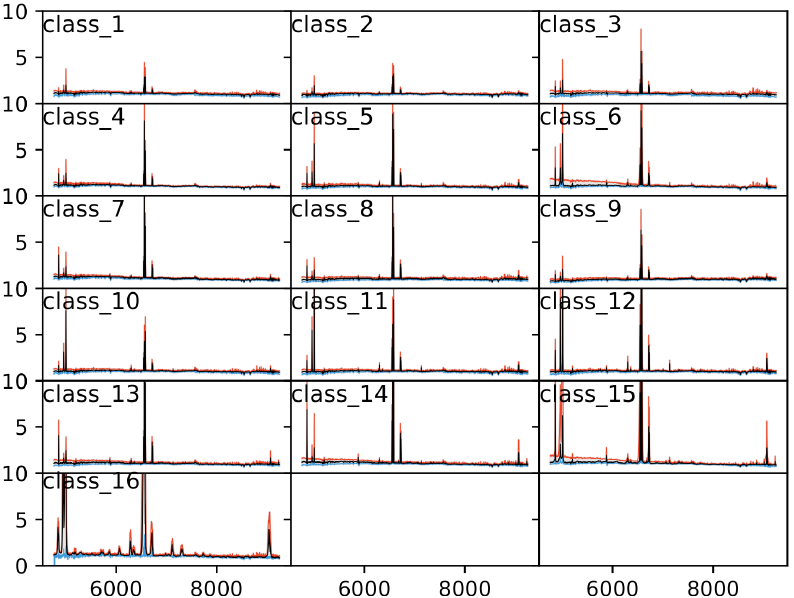}
        \caption{Median spectra (black) of every class for NGC~1068 with their dispersion (10 and 90\% quantiles in blue and red). The y-axis is in arbitrary units with the same scale for all plots.}
        \label{Fig:NGC1068_AllMedSpec}
\end{figure*}

\begin{figure*}
        \includegraphics[width=\linewidth]{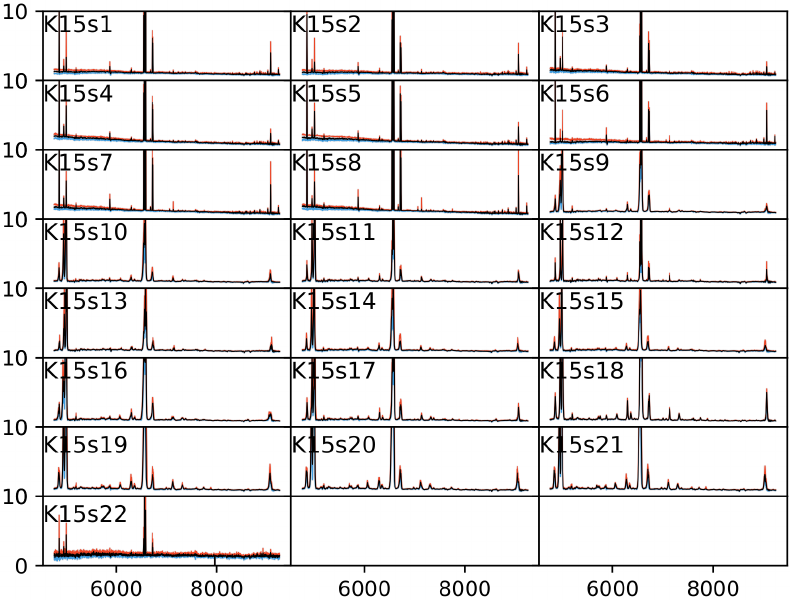}
        \caption{Median spectra (black) of every subclass for NGC~1068 subclassifcation of class 15 with their dispersion (10 and 90\% quantiles in blue and red). The y-axis is in arbitrary units with the same scale for all plots.}
        \label{Fig:NGC1068_AllMedSpecSubclassif}
\end{figure*}

In this section, we show in Fig.~\ref{Fig:NGC1068_AllMedSpec} the spectra of the classes from the classification of NGC~1068 as described in Sect.~\ref{NGC1068_Results_GeneralStudy}, and in Fig.~\ref{Fig:NGC1068_AllMedSpecSubclassif} the spectra of the classes from the subclassification of class 15 as described in Sect.~\ref{NGC1068_Results_Nucleus}.

\section{Analysis of unaligned spectra in NGC~1068}
\label{Ap:NGC1068undereshifted}

\begin{figure}
        \includegraphics[width=\linewidth]{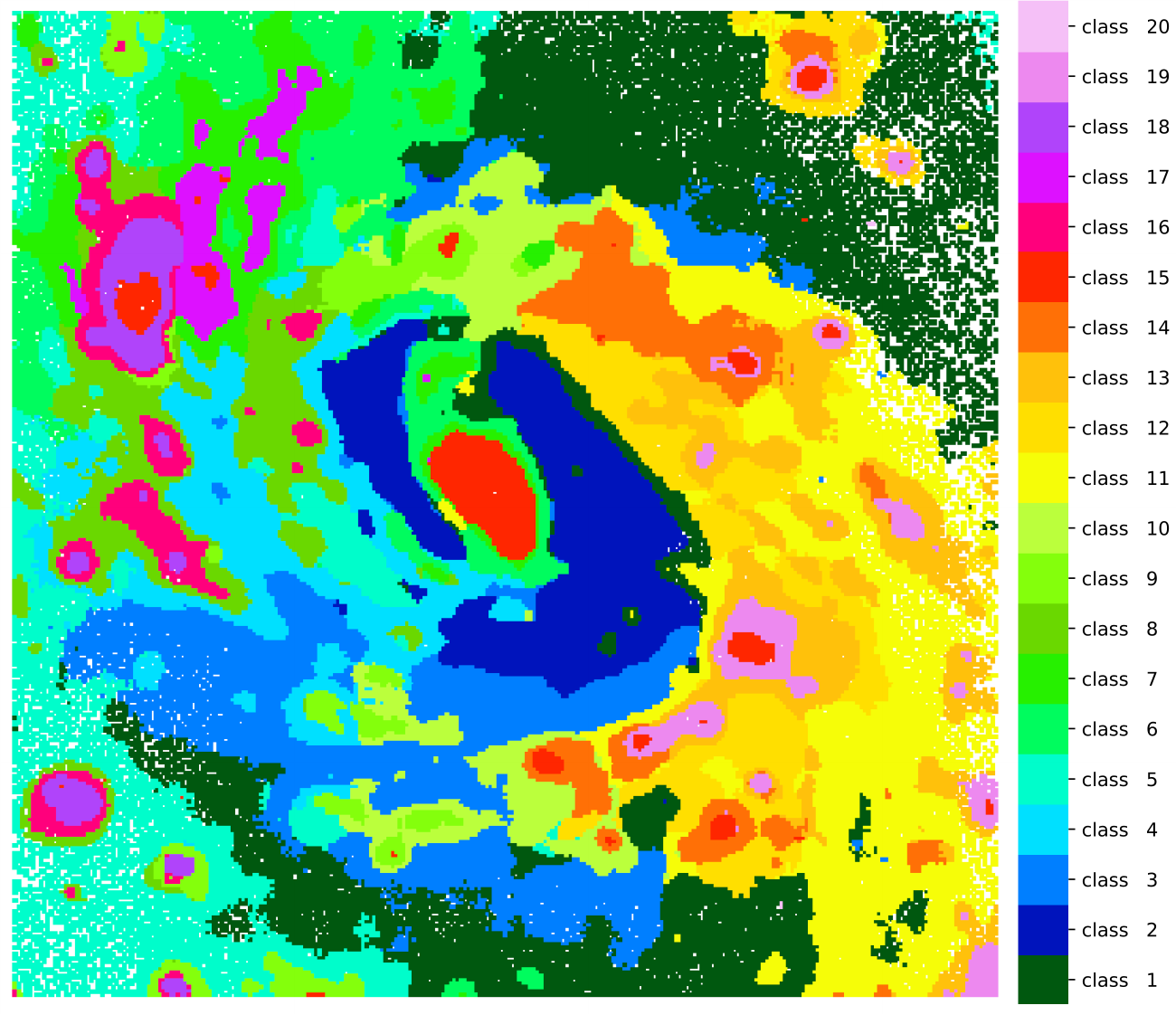}
        \caption{Class map of NGC~1068 with unaligned spectra. To be compared with Fig.~\ref{Fig:NGC1068ClassmapK16}.}
        \label{Fig:class_image_K20_NoDeredshifted}
\end{figure}

\begin{figure}
        \includegraphics[width=\linewidth]{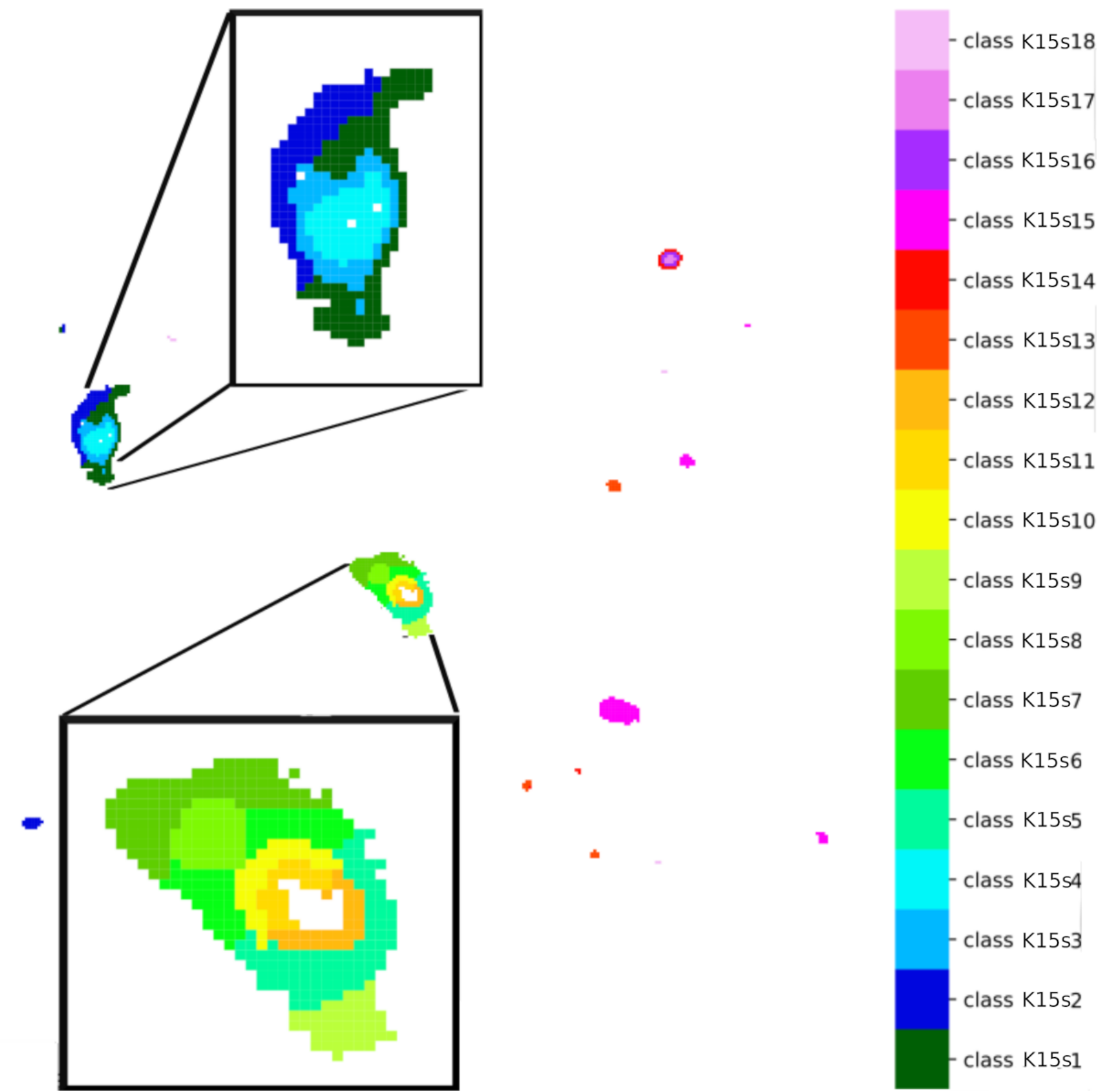}
        \caption{Class map of NGC~1068 subclassification of class 15 with its unaligned spectra. To be compared with Fig.~\ref{Fig:NGC1068BptDiagK16}.}
        \label{Fig:class_image_K18_NoDeredshifted}
\end{figure}

When comparing spectra in a clustering or classification context, correcting for the Doppler effect is essential. We included this here automatically with the algorithm ALFA. This process may fail when the emission lines are multiple or enlarged by small-scale motions. It may then be interesting to perform the unsupervised classification without this correction and hope that the algorithm, here Fisher-EM, will take the relative kinematics into account in addition to the physical differences. We illustrate this idea by repeating the classifications made on NGC~1068, but without an a priori alignment of the spectra, first, on the whole data cubes, and second, on the subclassification of class 15.

For the whole data cubes, the optimum is obtained for 20 classes. The class map (Fig.~\ref{Fig:class_image_K20_NoDeredshifted}) shows a clear distinction between the blue- and redshifted sides of the galaxy. The kinematics clearly dominates the classification, and despite a nearly identical number of classes (20 vs 22; Sect.~\ref{NGC1068_Results_GeneralStudy}), intrinsically similar spectra are less well identified than with redshift correction (Fig.~\ref{Fig:NGC1068ClassmapK16}). For instance, the ring is much less prominent, and the regions impacted by the jet are not emphasised. We conclude that the unsupervised classification without aligning the spectra might provide a first guess of the kinematics structure in the data cubes, but does not provide a sufficiently discriminating classification for the physics.

The subclassification of class 15 with the unaligned spectra yields an optimum of 18 classes, again nearly identical to the number (16) obtained with aligned spectra (Sect.~\ref{NGC1068_Results_Nucleus}). The separation between the nucleus and the other regions is also perfect, but different classes generated by the kinematics between the E and the W side of the galaxies are visible (Fig.~\ref{Fig:class_image_K18_NoDeredshifted}). We also find the same structures in the nucleus, with two roundish structures and the same asymmetries NE versus SW. The only differences in the nucleus is the much smoother class map than when we aligned spectra. This could show that there might be a consistent kinematics at the scale of the nuclear region, but this is probably more complex at the spaxel level.

\section{Analysis of NGC~4151}
\label{Ap:NGC4151_IndSubclassHddcFem}

\subsection{Fisher-EM}

\begin{figure}
        \includegraphics[height=0.9\textheight]{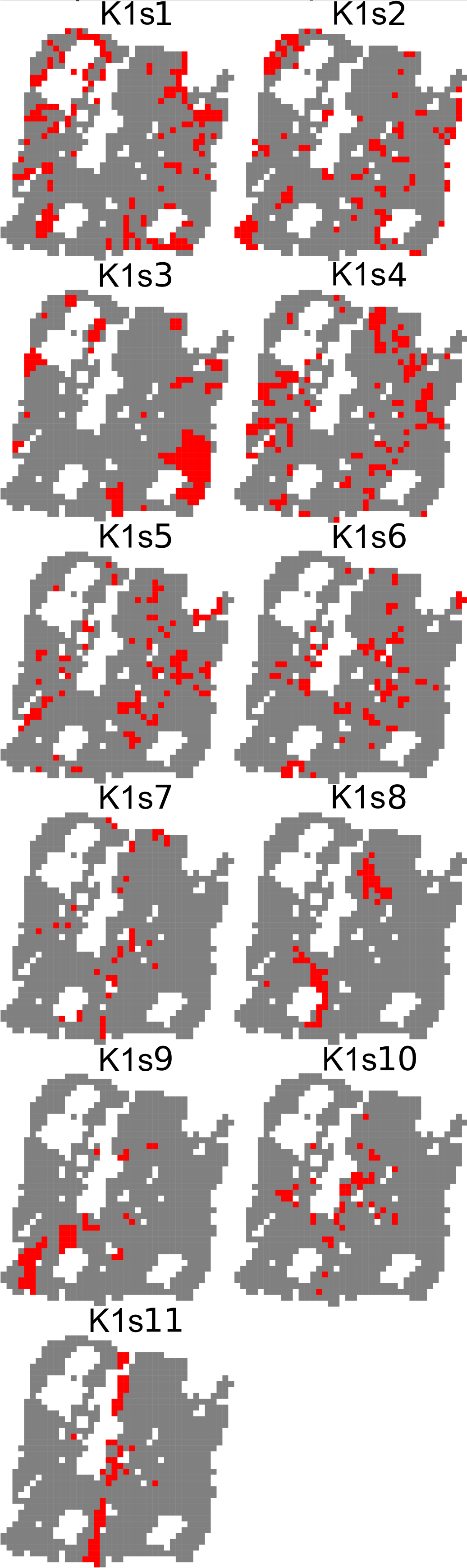}
        \caption{Maps of individual subclasses of NGC~4151 class 1 with Fisher-EM. Each subclass is highlighted in red, and the others are plotted in grey. See Fig.~\ref{Fig:NGC4151ClassmapSummaryFEM} and Sect.~\ref{NGC4151:resultsFEM}.}
        \label{Fig:NGC4151_IndSubclassFem}
\end{figure}

For a better readibility of the classification results, we show a map for each individual subclass of NGC~4151 with Fisher-EM (Fig.~\ref{Fig:NGC4151_IndSubclassFem}).

 \subsection{Analysis with the HDDC method}
\label{App:HDDC}

For comparison with the results of Fisher-EM (see Sect.~\ref{NGC4151_discussion}), we used another algorithm, called high-dimensional data clustering (HDDC) algorithm, which fits the data in class-specific subspaces \citep{Bouveyron2007}. For this reason, we might expect that a subclassification is not generally necessary. Except for an illustration on hyperspectral images of Mars made in this latter paper, this is the first use of HDDC in astrophysics. This algorithm is available in the \R\footnote{\url{https://www.r-project.org/}}\ package HDClassif \citep[command \textit{hddc},][]{Bouveyron2007}.

\subsubsection{High-dimensional data clustering algorithm}
\label{Ap:method:HDDC}

Here, the GMM is applied directly to the data themselves,
   \begin{equation}
       \centering
       Y|_{Z=k} \sim \mathcal{N}(\mu_k, \Sigma_k)
       \label{eq:XHDDC}.
   \end{equation}
\citet{Bouveyron2007} proposed to constrain the GMM through the eigen-decomposition of the covariance matrix $\Sigma_k$.
Let $Q_{k}$ be the orthogonal matrix with the eigenvectors of
$\Sigma_{k}$ as columns.
The class conditional covariance matrix $\Delta_{k}$ is therefore
defined in the eigenspace of $\Sigma_{k}$ by 
\begin{equation}
\Delta_{k}=Q_{k}^{t}\,\Sigma_{k}\, Q_{k}.
\label{eq:Definition-Delta-i}
\end{equation}
 The matrix $\Delta_{k}$ is thus a diagonal matrix that contains
the eigenvalues of $\Sigma_{k}$. It is further assumed that $\Delta_{k}$ is divided into two blocks, 
\begin{equation}
\small \Delta_k=\left(  \begin{array}{c@{}c} \begin{array}{|ccc|}\hline a_{k1} & & 0\\ & \ddots & \\ 0 & & a_{kd_k}\\ \hline \end{array} & \mathbf{0}\\ \mathbf{0} &  \begin{array}{|cccc|}\hline b_k & & & 0\\ & \ddots & &\\  & & \ddots &\\ 0 & & & b_k\\ \hline \end{array} \end{array}\right)  \begin{array}{cc} \left.\begin{array}{c} \\\\\\\end{array}\right\}  & d_{k}\vspace{1.5ex}\\ \left.\begin{array}{c} \\\\\\\\\end{array}\right\}  & (p-d_{k})\end{array}
\label{eqHDDC:Delta},
\end{equation}
with $a_{kj}>b_{k}$, $j=1,...,d_{k}$. The parameter $d_k\in\{1,\dots,p-1\}$ is unknown and can be considered as the dimension of the subspace associated to the kth group, that is, the number of dimensions required to describe the main features of this group. The second block is assumed to represent the noise described by $b_{k}$.

Because of the class-specific subspaces, the number of parameters to optimise is large. By fixing some parameters to be common within or between classes,
\citet{Bouveyron2007} proposed several parsimonious models that correspond to different regularisations. The estimation of all the parameters uses a classical EM algorithm.

\subsubsection{Application of HDDC to NGC~4151}
\label{Ap:NGC4151}

\begin{figure}
        \includegraphics[width=\linewidth]{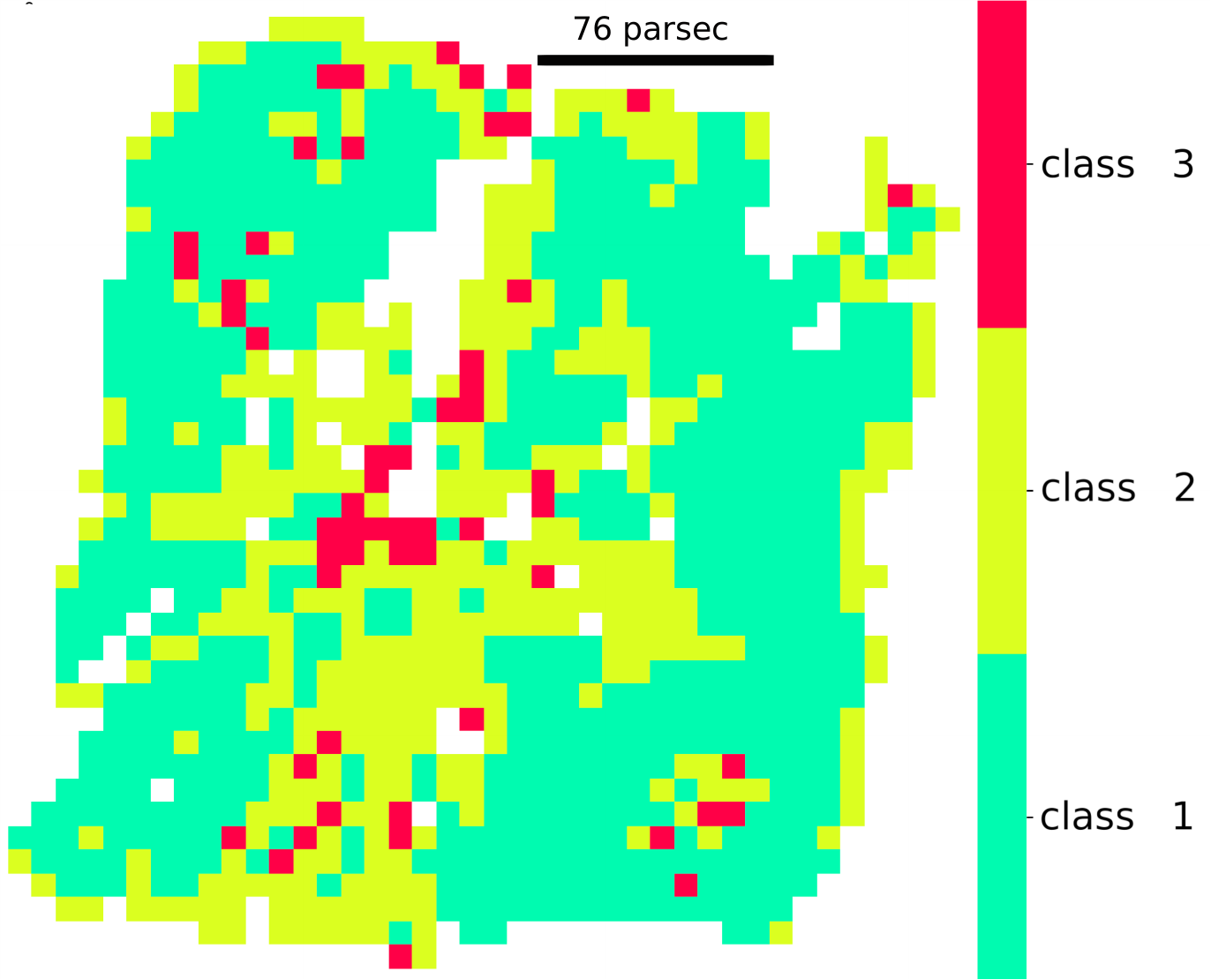}
        \caption{Class map for NGC~4151 with three classes with HDDC. The class nomenclature is explained in Fig.~\ref{Fig:JKB18classmap}.}
        \label{Fig:NGC4151ClassmapHDDC}
\end{figure}

\begin{figure}
        \includegraphics[width=\linewidth]{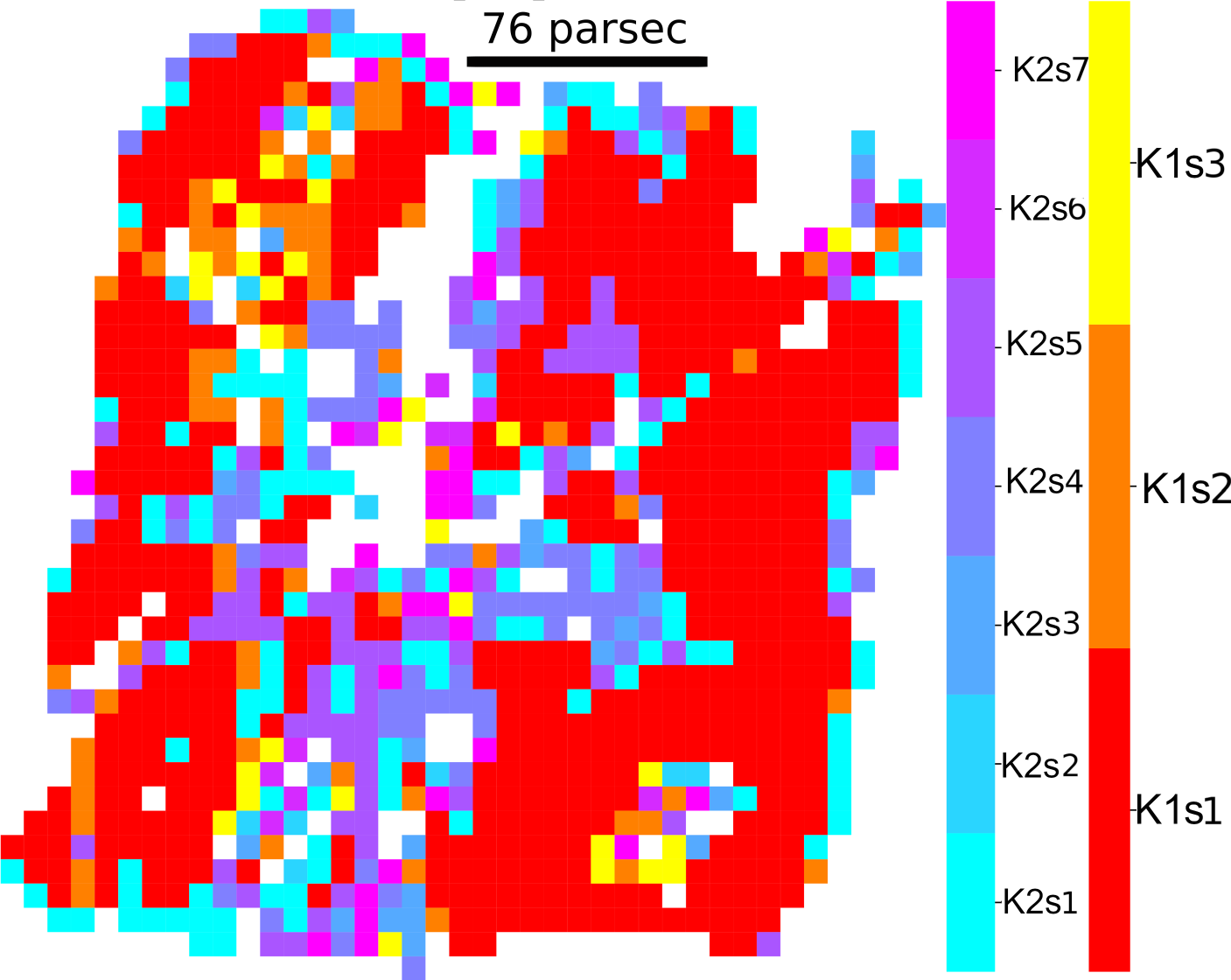}
        \caption{Class map summary for NGC~4151 for the subclassifications of classes 1 (reddish colours) and 2 (blueish colours) of Fig.~\ref{Fig:NGC4151ClassmapHDDC} with HDDC. The class nomenclature is explained in Fig.~\ref{Fig:JKB18classmap}.}
        \label{Fig:NGC4151ClassmapSummaryHDDC}
\end{figure}

\begin{figure}
        \includegraphics[height=0.9\textheight]{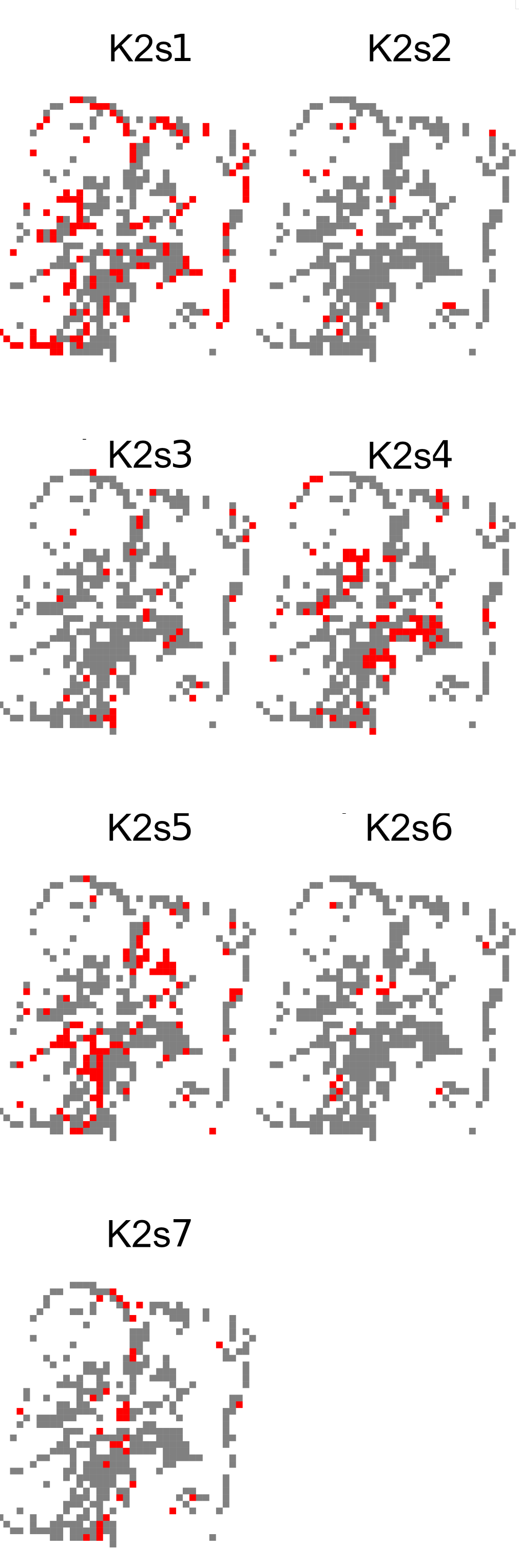}
        \caption{Maps of individual subclasses of NGC~4151 class 2 with HDDC. Each subclass is highlighted in red, and the others are plotted in grey. See Fig.~\ref{Fig:NGC4151ClassmapSummaryHDDC}.}
        \label{Fig:NGC4151_IndSubclassHddc}
\end{figure}

The classification of NGC4151 with HDDC led to an optimum of three classes (Fig.~\ref{Fig:NGC4151ClassmapHDDC}). Classes 1 and 2 represent the majority of spaxels. Class 2 encompasses subclasses K1s4 to K1s11 except for K1s9 of the Fisher-EM subclassification. Class 1 corresponds to subclasses K1s1 to K1s3 and to classes 2 to 9 of the Fisher-EM classification. Because classes 1 and 2 of the HDDC classification are large and show high dispersion, we subclassified each of these classes (Fig.~\ref{Fig:NGC4151ClassmapSummaryHDDC}. For better visibility, we built maps of individual classes (Fig.~\ref{Fig:NGC4151_IndSubclassHddc}). Both subclassifications highlight several structures similar to those obtained by Fisher-EM both spatially and physically, but they appear to be less precise. For instance, the SE structure of the Fisher-EM subclass K1s9 is not found by HDDC. In addition, the HDDC classes are more dispersed in their spectra than Fisher-EM. 

Hence, we conclude that the HDDC algorithm does not avoid the necessity of subclassification and does not bring more information than Fisher-EM for our data for NGC~4151. However, both algorithms yield consistent results, which means that our analyses are more robust.

\end{appendix}
%


\end{document}